\documentclass{article}

\usepackage[english]{babel}

\usepackage[letterpaper,top=2cm,bottom=2cm,left=3cm,right=3cm,marginparwidth=1.75cm]{geometry}
\usepackage{cite}
\usepackage{comment}
\usepackage{amsmath}
\usepackage{graphicx}
\usepackage{xcolor}
\usepackage[colorlinks=true, allcolors=blue]{hyperref}

\title{50 Years of SUSY and SUGRA, circa 1974-2024, and Future Prospects}

\author{Pran Nath\footnote{\href{mailto:p.nath@northeastern.edu}{p.nath@northeastern.edu}}
 \\~\\
\textit{\normalsize Department of Physics, Northeastern University,} \\
\textit{\normalsize 111 Forsyth Street, Boston, MA 02115-5000, U.S.A.} \\
}

\begin{document}
\maketitle
\tableofcontents

\begin{abstract}
The development in the early seventies of supersymmetry, in the mid-seventies of gauge supersymmetry and supergravity, and in the early eighties of gravity mediated breaking of supersymmetry and of supergravity grand unification  have led to remarkable progress in the pursuit of unification of fundamental interactions of particle physics. They have also led to the  intertwining of particle physics, cosmology, and strings. Since supersymmetry and supergravity are manifest in the low energy limit of superstring below the Planck scale, experimental test of them are of interest regarding the validity of the superstring itself.
For that reason, over the past decades, after the advent of supersymmetry and SUGRA models, there have been sustained experimental searches for supersymmetry at colliders, in precision experiments, and in astrophysical and cosmological data.
The SUSY and SUGRA models have also had deep impact on theories related to inflation, dark matter, and dark energy. The purpose of this article is to provide a view from the bridge of these developments over the past fifty years circa 1974-2024.\\


\end{abstract}

\section{Early history of SUSY}

 The first papers in supersymmetry appeared in the period 1971-1974. Thus in 1971 Yu. A. 
 Golfand and E.~P.~ Lyktman \cite{Golfand:1971iw} 
 extended the Poincaré algebra to include spinor generators  $Q_\alpha, \bar Q_{\dot \beta}$ which
 in the two component notation satisfy the relation
 
\begin{equation*}
\{ Q_\alpha, \bar Q_{\dot\beta} \} = 2 \sigma^\mu_{\alpha\dot\beta} P_\mu,~~~
\{Q_\alpha, Q_\beta\}=0,
\end{equation*}
and the relations $[P_\mu, Q_\alpha] = 0$, $ [M_{\mu\nu}, Q_\alpha]= -\frac{i}{2} (\sigma_{\mu\nu})_\alpha{}^\beta Q_\beta$ along with their h.c. relations.
Around the same time Ramond\cite{Ramond:1971gb}
discovered a graded algebra on the worldsheet generated by the fermionic modes $F_r$ which extend the algebra of the Virasoro 
generators $L_n$\cite{Virasoro1970}, given by  $[L_m, L_n]
= (m-n) L_{m+n}
+ \frac{c}{12} m(m^2-1)\,\delta_{m+n,0}$, where $c$ is the central charge, to include the relations 
\begin{equation*}
\{ F_r , F_s \} =2 L_{r+s}+  \frac{c}{3}\left(m^2 - \frac{1}{4}\right)\delta_{m+n,0}, ~~[L_m, F_r]=(\frac{m}{2} -r) F_{m+r}.
\end{equation*}
Ramond's construction was in two dimensions, which eventually led to spacetime supersymmetry in string theory. In 1973
Volkov and Akulov \cite{Volkov:1973ix}
constructed the first model of spontaneous SUSY breaking. In 1974
Wess and Zumino \cite{Wess:1973kz,Wess:1974tw}
developed the first  renormalizable 4D 
Lagrangian model invariant under supersymmetric field theory transformations. Subsequent to the work of Wess and Zumino, 
Salam and Strathdee introduced superspace to describe supersymmetry. They also formulated Lagrangians in superspace which were invariant under supergauge transformations\cite{Salam:1974yz,Salam:1974za}.The superspace consists of extending the ordinary bosonic spacetime given by $x^\mu$ with inclusion  of fermionic co-ordinates $\theta_\alpha$, which are Majorana spinors,  which satisfy the anti-commutation relations 
$\{\theta_\alpha,\theta_\beta\}=0$. One of the remarkable features of supersymmetry is that it overcomes the restriction of the Coleman-Mandula theorem\cite{Coleman:1967ad} which does not 
allow unification of space-time and internal symmetries within the conventional Lie algebra framework, which involves only commutators. This theorem is overcome within the graded lie algebra, allowing supersymmetry to be an allowed symmetry of the S-matrix\cite{Haag:1974qh}.


\section{Origin of Local Supersymmetry}
The first formulation of local supersymmetry is Gauge Supersymmetry \cite{Nath:1975nj,Arnowitt:1975xg},
which is based  on the tangent space supergroup
$\mathrm{OSp}(3,1|4N)$.
 {The $O(3,1)$ part of the group refers to Lorentz symmetry of spacetime  and the
 $Sp(4N)$ part is a symmetry group that acts on the fermionic co-ordinates while the bar indicates that it is a supergroup  with bosonic and fermionic generators which form its algebra. This group is often referred to as orthosymplectic. The gauge theory with the tangent space group $\mathrm{OSp}(3,1|4N)$
   includes Einstein gravity and local supersymmetry}. Spontaneous breaking of gauge supersymmetry
leads to a vacuum state which is globally supersymmetric and is given by the supermetric
$g_{AB} =\text{diag} (\eta_{ab},  k\eta_{\alpha\beta}\,\delta_{ij})$,
where $k$ is a constant that enters along with the metric in the fermionic line element. The gauge group of gauge supersymmetry
is the largest supergroup one can have which means that any other 
theory of local supersymmetry must be a subpiece of gauge supersymmety. Thus after the formulation of gauge supersymmetry in 1975, the current standard supergravity was formulated in
1976\cite{Freedman:1976xh,Deser:1976eh} which has the tangent space group $O(3,1)\times O(N)$
which is a subpiece of the $OSp(3,1|4N)$. 
The above leads one to expect that the dynamics of standard supergravity 
theory would also arise from the dynamics of gauge supersymmetry
 in a  certain limit. In fact, this turns out to be the case
 as shown in several works\cite{Nath:1976ci,Arnowitt:1978jq,Nath:1979yn}. 
Geometrically, the connection between gauge supersymmetry and supergravity is the following:
One can produce 
the standard supergravity with the tangent space group $O(3,1)\times O(N)$ from gauge supersymmetry by including only the spin 2 and spin $3/2$ fields
and letting $k\to 0$ to reduce the Riemannian geometry of gauge supersymmetry to the non-Riemannian superspace geometry of the standard supergravity theory.
The contraction produces the desired torsions ~\cite{Nath:1979yn}
needed in  the superspace formulation of supergravity
~\cite{Wess:1977fn,Brink:1978iv,Brink:1978sz} and  the tangent space group  $OSp(3,1|4N)$ 
of gauge supersymmetry  reduces ~\cite{Nath:1979yn} to the tangent space group  $O(3,1)\times O(N)$  of supergravity.
Further, the equations of motion of supergravity emerge from the
dynamical equation of gauge supersymmetry with the particle content of  the spin 2 and spin 3/2 fields .
Thus, the $k\to 0$ limit of the gauge supersymmetry geometry correctly produces the  supergravity geometry in the superspace. The geometry contraction of gauge supersymmetry to supergravity may be viewed as the counterpart of Inonu-Wigner\cite{Inonu:1953sp} group contraction.
A more detailed discussion of the limit $k\to 0$ of gauge supersymmetry to the standard supergravity is given in Appendix A.
We note here that, after the work 
of \cite{Freedman:1976xh,Deser:1976eh} there were different formulations of supergravity including \cite{
AHCThesis,Chamseddine:1976bf,MacDowell:1977jt,Siegel:1978mj} among several others.
\section{Supergravity grand unification: SUGRA}
Supersymmetry came to prominence in particle physics due to its solution to the gauge hierarchy problem~\cite{hierarchy}.
In the standard model loop corrections to the Higgs boson mass$^2$ receive a  correction
from the quarks/leptons in the loop which is proportional to  $\Lambda^2$, where $\Lambda$ is the cut-off which can be as large as the Planck scale 
$M_{Pl}=(8\pi G_N)^{-1/2}$ where $G_N$ is Newton's constant, and numerically
$M_{Pl}\equiv \kappa^{-1}\sim 2.4\times 10^{18}$ GeV. This requires a huge amount of fine tuning to bring the Higgs boson mass down to the electroweak scale.
 In a supersymmetric theory, this large correction is  canceled by squark/slepton loops~\cite{hierarchy,Kaul:1981wp}. 
 A further discussion of this phenomena and its relation to no-renormalization theorems is discussed in Appendix B.
 However, an exact supersymmetry is not viable since supersymmetric particles are not yet observed.
   Thus, to achieve a viable model, one needs sparticles to be massive enough to have evaded detection.    
The simplest possibility is to break supersymmetry spontaneously. However, it is 
difficult to achieve a  viable breaking for global
supersymmetry.
One  reason for the difficulty is due to the Hamiltonian for a supersymmetric model being positive semi-definite which means that the lowest energy state is
   the state of unbroken supersymmetry with $E=0$ and states arising from breaking of supersymmetry lie above.
   Another issue  concerning spontaneous breaking of supersymmetry  relates to 
   the emergence of a massless Goldstino field,
   which is in conflict with observation. Phenomenologically, one may add 
   soft terms that leave the ultraviolet behavior of the theory 
   unchanged~\cite{Girardello:1982pd}. The addition of soft terms allows 
for the grand unification of the  gauge couplings $SU(3)\times SU(2)\times U(1)$~\cite{Dimopoulos:1981yj}. These analyses provide further motivation for 
   theoretically generating soft terms within a fundamental particle physics.

 \subsection{Applied N=1 supergravity: couplings to matter and gauge multiplets}
Due to the difficulties  in global SUSY models in achieving acceptable models
    where spontaneous breaking gives rise to soft terms, it was natural then to turn to 
    formulations of grand unification within the framework of local supersymmetry.
    One of the first items needed to construct a supergravity  unified model is to obtain the coupling of supergravity to matter multiplets which include chiral superfields which would describe quarks and leptons and Higgs, and vector  superfields which would describe gauge bosons for the  gauge group. However, in 1981 coupling of supergravity to only one chiral field was known through the work of Cremmer et al.~\cite{Cremmer:1978hn} who used the rules of  super-conformal
 tensor calculus~\cite{Ferrara:1978jt,Stelle:1978ye}.
             Thus, one needed to extend the coupling to multiple chiral fields and to include coupling also to the gauge fields. Such constructions were later labeled applied supergravity. One of the remarkable aspects of the analysis is that the couplings of supergravity to matter are determined in terms of  three  functions: the superpotential $W(Z_A)$, the 
 K\"ahler potential $K(Z_A, Z_A^\dagger)$ and a gauge kinetic function
$f_{AB}(Z_A)$  where $Z_A$ are the bose components of the chiral multiplets~\cite{Chamseddine:1982jx,Cremmer:1982wb,Cremmer:1982en,appliedsugra}.
The scalar potential that emerges from the analysis takes the form
\begin{align}
 V=-\frac{1}{\kappa^4}e^{-G}(3+(G^{-1})^A_BG_{,A}G,^{B})
+\frac{1}{2\kappa^4}
Re(f^{-1})_{rs} D^r D^s
\label{3}
\end{align}
where $G$ is defined by
\begin{align}
G=-\kappa^2 K(Z,Z^{\dagger}) -ln(\frac{\kappa^6}{4}|W|^2).
\label{1} 
\end{align} 
In the above, $(G^{-1})^A_B$ is the inverse of $G_{,A}^{~B}$ and  $D^r = -g_r \kappa^{-2} (G_i(T^r)_{ij} Z_j)$, {where $g_r$ is the gauge coupling constant for the gauge group with generators $T^r$.}
The full potential contains all the chiral fields and the full Lagrangian involves the gauge kinetic  function $f_{AB}$ which enters prominently in fermionic gauge interactions. The global supersymmetric limit of the potential can be obtained by letting {$M_{Pl} \rightarrow \infty$, and} correspondingly $\kappa\to 0$.\\

 We discuss now the important distinctive feature of the supergravity scalar potential relative 
 to the  potential from the global SUSY case. Thus, while the scalar potential was positive definite for global SUSY, this is not the case for the supergravity scalar potential. We illustrate the distinction for the case of  $SU(5)$ grand unification where the GUT symmetry is broken by the 24-plet of Higgs $\Sigma_x^{~y}$ with the superpotential $W=\frac{1}{3} \Sigma^3+\frac{1}{2}M \Sigma^2$. In this case the potential has a minimum for one of the following three cases: (i)$\Sigma_y^{~x}=0$; (ii) $\Sigma_y^{~x}=\frac{1}{3} M[\delta_y^{~x}-5 \delta_5^{~x}\delta_y^{~5}]$; (iii) 
$\Sigma_y^{~x} =M[2 \delta_y^{~x} - 5(\delta_5^{~x}\delta_y^{~5}+ \delta_4^{~x}\delta_y^{~4})]$.
The three cases here describe the symmetry breaking patterns for the field $\Sigma$. Thus, (i) corresponds to no symmetry breaking, (ii) describes the breaking of $SU(5)$ to $SU(4)\times U(1)$,  and (iii) describes the breaking of $SU(5)$ to the standard model gauge group $SU(3)_C\times SU(2)\times U(1)$. All these vacua are degenerate for the case of global supersymmetry. However,  for the case of supergravity potential  the degeneracy is lifted, and the lifting is proportional to $\kappa^2 W(\Sigma_0)$ (see\cite{Chamseddine:1982jx,ovrut,Weinberg:1982id}). Further, if one chooses solution (iii) to be the physical vacuum and adjusts the vacuum energy to vanish for this case by adding a constant in the superpotential, then the other two vacuum states have lower vacuum energies. Weinberg\cite{Weinberg:1982id} has argued that the flat vacuum for case (iii) would be stable against finite perturbations even though it is not the lowest energy state. In the analysis so far, while the gauge symmetry of the gauge group is broken, supersymmetry is still intact. In order to generate a viable model one needs to break supersymmetry. 

\subsection{SUSY breaking}
 In addition to lifting the degeneracy of vacuum states in the breaking of the GUT symmetry, super gravitational interactions help break supersymmetry spontaneously via the 
  super Higgs phenomenon in a viable fashion. This becomes  possible  because the  scalar potential $V$ is no longer positive  semidefinite. A simple demonstration of the super Higgs mechanism is given by assuming a linear 
 $W = m^2(Z + B)$ where Z is a singlet super Higgs field. Using this superpotential in 
 the supergravity potential $V$ produces a VEV for  $<Z>$ of the Planck mass size so that
$\kappa <Z> = O(1)$\cite{polonyi}. Furthermore, the parameter $B$ allows one to adjust the cosmological constant to zero.  More generally, one can assume a $W$ of the form 
$W = m^2f(\kappa Z)/\kappa$ that would again lead to a minimum of the potential 
  when $\kappa <Z> = O(1)$ that gives  $<W> =O(m^2/\kappa)$. After breaking, the gravitino
   develops a mass of size $\kappa^2<W> = \kappa m^2$. Further, the super Higgs breaking
   solves the problem of global supersymmetry where spontaneous breaking of 
   supersymmetry leads to a massless goldstino. In the super Higgs breaking
   using the supergravity potential $V$, the goldstino is absorbed by the massless spin
   $3/2$ gravitino to become 
   massive\cite{volkov,Cremmer:1978hn}.  \\ 

  In order to set up supergravity grand unification we need to make sure that soft terms
  with masses of the order of Planck mass or GUT mass $M_G$ do not arise. Thus, let
  $Z_A$ stand for all fields that enter the analysis 
  so that $Z_A=(Z_a, Z)$, where $Z_a$ are all fields in the visible sector, and $Z$ 
  is the super Higgs field. In this case, couplings of type $Z_a Z_b Z$ in the superpotential
  would generate Planck scale masses for the visible sector fields. To avoid this 
  undesirable result, we must have another sector, i.e., a hidden sector where $Z$ resides, 
  and we assume that there is no contact between the visible and the hidden sectors at the level of
  superpotential so that $W_t(Z_a, Z)=W (Z_a)+ W_h(Z)$ where $W_t$ is {the} total superpotential,
  $W(Z_a)$ is the superpotential for the set of fields in the visible sector, and $W_h(z)$ is the superpotential for the hidden sector.  Separation of the two sectors guaranties that communication between the two sectors would be suppressed by factors of $1/M_{Pl}$. This works when all the 
  fields in the visible sector are light and one finds that the soft terms are typically of size
  $m_s=\kappa m^2$, which can lie in the electroweak region when $m\sim 10^{11}$ GeV.
  However, for grand unified theories one has both light and heavy sectors, so that 
  $Z_a=(Z_\alpha, Z_i)$ where $Z_\alpha$ are light and 
  $Z_i$ are heavy. In this case,
 mass squared corrections can become large, and one may have terms of the type

 \begin{align}
m_s M_G, ~m_s M_G (\kappa M_G), \cdots,  m_s M_G (\kappa M_G)^k,
\label{msmG}
   \end{align} 
  which can upset the gauge hierarchy for $k\leq 6$. In\cite{Nath:1983aw} it  was shown that the 
  imposition of the following constraints  
 
\begin{align}
  W_{,\alpha i} = O(\kappa  m^2);  ~~W_{,\alpha \beta} = O(\kappa m^2)
  \label{alphai}
\end{align}
are sufficient to protect the low energy physics from GUT mass scale corrections.
 Using the above,  a remarkable form for the effective low energy theory arises. 
 Thus, for the case where  $f_{\alpha\beta} = \delta_{\alpha\beta}$  
 and the Kahler  potential 
 is taken to be flavor independent, one finds 
\begin{align}
V_{eff} = |\frac{\partial \tilde W}{\partial Z_{\alpha}}|^2  +
m_0^2 Z_{\alpha}^{\dagger} Z_{\alpha} + (B_0 W^{(2)} + A_0 W^{(3)} + h.c)  + \frac{1}{2}[g_\sigma
\kappa^{-2}
(G_{\alpha}
(T^{\sigma} Z)_{\alpha})]^2,
\label{soft}
      \end{align}
where $\tilde W$ is the superpotential that contains only
quadratic and cubic functions of
 the light fields, i.e., $\tilde W(Z_{\alpha})
= W^{(2)}(Z_{\alpha}) + W^{(3)}(Z_{\alpha})$, 
and $m_0, A_0, B_0$ are soft breaking
parameters of size $m_s$. 
 For the case of the minimal supersymmetric standard model $W^{(2)}$ would contain a term of the form $\mu H_1H_2$ where $\mu$ has the dimensions of mass. A term of this type can arise
 from a cubic interaction $SH_1H_1$. It turns out that $S$ can develop a VEV of size $m_s$\cite{Chamseddine:1982jx} that produces a contribution of the electroweak size.
 A more convenient mechanism is to assume a term in the Kahler  potential of the form $c H_1H_2+ h.c$, and then make a Kahler  transformation to move it to the superpotential\cite{soniweldon,gm}.
 {Yet another solution to generate an electroweak size $\mu$ is  via breaking of the Peccei-Quinn (PQ) symmetry.
Thus suppose $H_1H_2$ in not invariant under the global $U(1)_{\mathrm{PQ}}$ symmetry but a higher 
dimensional operator $\frac{S^2}{M_{Pl}} H_1H_2$ is invariant due to $S$ carrying an appropriate PQ
charge. In this case while $\mu=0$ at the tree level, it is generated after spontaneous breaking of PQ.
Thus after spontaneous breaking of the PQ symmetry, the field $S$ develops a VEV so that
$<S>\sim f_a$ where $f_a$ is the axion decay constant and an $f_a\sim (10^{10}-10^{11})$ GeV can generate 
a $\mu$ of electroweak size\cite{Kim:1983dt}.
Another relevant reference relates to the reduction of a 10-d spacetime without generating R parity violating operators and a solution to the strong CP problem with  an electroweak size $\mu$ term\cite{Baer:2018avn}.
}
Next, consider the coupling $y_{ij\alpha} Z_iZ_j Z_{\alpha}$.
This coupling is forbidden unless the VEVs of $Z_i$ and $Z_j$ vanish.  
    We note that in addition to the mass growth for the squarks and the sleptons implied by
    Eq.(\ref{soft}), gauginos can also assume masses. This can happen at the loop level by the exchange of 
       heavy fields~\cite{can5,AlvarezGaume:1983gj,il-1,il-2,ilm,lajolla}. 
  In general, the gaugino mass term will have the form 
  $(\tilde m)_{\alpha\beta} \lambda^\alpha \gamma^0 \lambda^\beta$.
  A flat gauge kinetic term with exchange of heavy 
       fields gives rise to gaugino masses after SUSY breaking, 
       so that 
       
   \begin{align}
\tilde m_i= \frac{\alpha_i}{4\pi} m_{1/2} \frac{D(R)}{D(A)}
   \end{align}
       where $i$ refers to the gauge groups $SU(3)_C$, $SU(2)$, 
       and $U(1)$, $\alpha_i$
are the corresponding gauge coupling constants, $D(R)$ is the dimensionality of the exchanged representation, and $D(A)$ is the dimensionality of the adjoint representation.
          Alternatively, they can 
     be generated more directly using appropriate field dependence in the gauge kinetic energy 
     function\cite{Cremmer:1982en,Drees:1985bx}. 
     In general, the gaugino masses can  
     all be different consistent with $SU(3)_C\times SU(2)\times U(1)$ gauge invariance, i.e.,
     $\tilde{m}_i, i=1,2,3$, but the simplest possibility is that they are all equal at the unification scale
     with $\tilde{m}_i=m_{1/2}$. 
       We note here several interesting works subsequent to the work of  \cite{Chamseddine:1982jx}, chief among them  are ~\cite{Barbieri:1982eh,Ibanez:1982ee,Hall:1983iz}. 
       For a historical perspective on the SUGRA unification, see 
       \cite{Chamseddine:2000nk,Chamseddine:2004ty,Arnowitt:2012gc,Nath:2015ska}. Further work on soft terms within  
    supergravity and strings can be found in  
\cite{Kaplunovsky:1993rd,Brignole:1993dj,Camara:2003ku,Grana:2003ek} and in intersecting D-brane models in 
    \cite{Kors:2003wf,Lust:2004fi}. For a more detailed discussion of supersymmetry breaking in fields, strings and branes see \cite{Dudas:2025ubq}.


\subsection{Radiative breaking of electroweak symmetry}
In a grand unified theory, gauge and Yukawa couplings are defined at the grand unification scale and evolved to the electroweak scale using renormalization group.
Their evolution in a  one- and two loop renormalization group order has been carried out in several works
\cite{Einhorn:1982pp,Machacek:1984zw, Machacek:1983fi,Machacek:1983tz,Machacek:1981ic}.
In supergravity grand unification, one needs in addition the evolution of the soft
parameters since they enter in the computation of the  squark, slepton, Higgs, 
higgsino, and gaugino masses. Their evolution up to two loops is carried out in
\cite{Inoue:1982ej,Inoue:1982pi,Jack:1994rk,Martin:1993zk,Martin:1993yx}.
In the standard model, breaking of the electroweak symmetry is
      ad hoc, where a  tachyonic mass term is added to the Lagrangian
      to accomplish the breaking. In global SUSY GUTs, $SU(2)_L \times U(1)$ symmetry breaking can occur as a consequence of radiative effects 
      arising from supersymmetry breaking\cite{Ibanez:1982fr}. As noted earlier  global supersymmetry has no fundamental way to achieve a 
      phenomenologically viable supersymmetry breaking. Since in a SUGRA model, supersymmetry can be broken spontaneously in a consistent fashion, it raises the possibility of breaking the electroweak symmetry from first principles.
      An elegant way to achieve this is through renormalization group effects and radiative breaking, which has been achieved in several works\cite{AlvarezGaume:1983gj,Ibanez:1984vq}.
      In SUGRA grand unified models breaking of the electroweak symmetry arises in a natural way consistent with charge and color conservation.
  {The Yukawa coupling of the top quark, with a mass of 175 GeV, plays a key role in inducing radiative breaking. Thus the  minimal SUGRA model involves a pair of Higgs doublets $H_i$ (i=1,2),
      where $H_2$ couples with the up quarks and $H_1$ couples with the down quarks and the leptons.    
      The RG equation for $H_2$ Higgs soft mass $m_{H_2}^2$ contains a dominant negative contribution from the top quark Yukawa coupling which drives $m_{H_2}^2$ negative at 
      low energy  leading to VEV formation 
      for the neutral components of $H_i$ so that $v_i=<H_i>$ which break the electroweak symmetry.} The VEV {$v= \sqrt{v^2_1+ v^2_2}$ }is determined
 to be $v\sim 246$ GeV by its relation to the 
 fermi constant, i.e. $G^{-1/2}_F=2^{1/4} v$.
 A convenient parameter in the analysis of low energy physics is $\tan\beta=<H_2>/<H_1>$, where
$<H_2>$ gives mass to the up quarks and
{$<H_1>$} gives mass to the bottom quarks and leptons. 
 
\subsection{mSUGRA }
 In Supergravity unification, the $\mu$ term arises as a consequence of SUSY breaking and  a scalar field getting a VEV which is the size of the
 SUSY breaking scale even though it is not a soft parameter. As noted earlier, $\mu$
 can also arise via a Kahler  transformation which transforms the term $cH_1H_2+ h.c.$
from the Kahler  potential to the superpotential. {Thus, the parameter space of 
Eq.(\ref{soft}) including the universal gaugino mass, and $\mu$ at the GUT scale, 
 is given by $m_0, m_{1/2}, A_0, B_0,\mu_0$. 
The renormalization group equations evolve these parameters from the GUT scale down to the
electroweak scale. Specifically $B_0,\mu_0$ become $B,\mu$ 
at the electroweak scale and minimization of the Higgs potential with respect to 
$v_1$ and $v_2$ gives us two equations, which at the tree-level read
\begin{equation}
 \mu^2=
 {m_{H_1}^2-m_{H_2}^2\tan^2\beta\over \tan^2\beta-1} -{1\over 2}M_Z^2, ~~ B\mu={1\over 2}
 \left(m_{H_1}^2+m_{H_2}^2+2\mu^2\right)\sin 2\beta .
\end{equation}
Using $M_Z$ as an input to fix the electroweak scale in the left hand equation above, one tunes $\mu$ reducing the parameter space from five to four. Further, one may use the right hand equation above to eliminate $B$ in favor of $\tan\beta$ and the residual  parameter space is then given by 
\begin{align}
m_0, m_{\frac{1}{2}}, 
A_0, \tan\beta, ~\text{sign}(\mu).
\label{msugra}
\end{align}
Here, we have 4 parameters and the sign of $\mu$ which arises because the minimization equations determine
$\mu^2$ leaving the sign of $\mu$ undetermined. The parameter space of the model given by Eq.(\ref{msugra}) 
  completely determines the sparticle phenomenology.} \\

An analysis of the sparticle mass spectrum within supergravity grand unification 
using the parameter set of Eq.(\ref{msugra}) was first done in \cite{Arnowitt:1992aq,Ross:1992tz}
where starting with the parameters of Eq.(\ref{msugra}) a computation of all sparticle masses was carried out. The model of Eq.(\ref{msugra}) has been variously called 
minimal supergravity model \cite{Barger:1993gh},  
mSUGRA\cite{Baer:1996kv}, and CMSSM\cite{Kane:1993td}.
The acronym mSUGRA appears most suitable since it points to its SUGRA origin,
and distinguishes it from
SUGRA models with non-universalities, which we label nSUGRA, which have also been extensively investigated in later studies.
   One may contrast the above, very restricted set of parameters,    
   with the global supersymmetric model MSSM which allows one to add a huge number of soft terms \cite{Dimopoulos:1995ju} consistent with the condition that the quadratic divergences cancel\cite{Grisaru:1979wc}.  
  So, without the benefit of a theory, regarding how soft SUSY parameters arise, one is left to constrain  
   over a hundred parameters to reduce them to just a handful. 
   However, as noted earlier, non-universalities can be added in a controlled fashion in non-universal SUGRA models.Thus, we may allow non-universalities in the Higgs boson sector, in the third generation sector consistent with FCNC, and in the gaugino sector where the gaugino masses $\tilde m_i$ (i=1,2,3)
   can acquire independent masses\cite{Matalliotakis:1994ft,Nath:1997qm,Corsetti:2000yq,Birkedal-Hansen:2002gkr,Baer:2000cb}.
   We note that in general the gaugino
   masses will have phases which enter in the electric dipole moments 
   of quarks and leptons(see,e.g., \cite{Nath:1997qm,Ibrahim:1997gj} and the references therein).
Although gravity mediated breaking of supersymmetry\cite{Chamseddine:1982jx,Barbieri:1982eh}, was the first consistent
mechanism of supersymmetry breaking to be discovered, several other
 mechanisms for supersymmetry breaking have appeared since then.
 Among them are gauge mediated\cite{Dine:1993yw},  anomaly mediated\cite{Giudice:1998xp,Randall:1998uk,Bagger:2000dh,Binetruy:2000md}, 
  gaugino mediated
 \cite{Dine:1995ag}, 
 moduli mediated\cite{Kachru:2003aw,Balasubramanian:2005zx},
modulus-anomaly  mediated {(generally known as mirage mediation)}\cite{Choi:2004sx,Choi:2005ge}
 dynamical breaking with a Green-Schwarz mechanism\cite{Arkani-Hamed:1998ufq}, and split supersymmetry\cite{Arkani-Hamed:2004zhs} among many others.\\
 
 {
We discuss now briefly the issue of naturalness as it relates to electroweak symmetry breaking
and generation of a little hierarchy in the sparticle mass spectrum. Thus for the universal boundary conditions at the GUT scale we  may write the electroweak symmetry breaking in terms of the GUT parameters $\mu_0,$ $m_0$, $m_{1/2}$, $A_0$
~\cite{Nath:1997qm,Chan:1997bi} so that
$\mu^2  = -\frac{1}{2}M_Z^2 +  m^2_0  C_1+ A^2_0 C_2
+ m^2_{1/2} C_3+ m_{1/2}
A_0 C_4$, where $C_i$ and $\mu$  are dependent on the renormalization group scale. If the $C_i$ are all positive, one would expect 
the soft parameter and $\mu$ to be order the electroweak scale. However, there are regions of the parameter space where $C_1$ can vanish and even turn negative which allows $m_0$ 
to get large, i.e.,  as large as $O(10)$ TeV. This is the hyperbolic branch of radiative breaking.
In this region the weak scale becomes insensitive to $m_0$ and a  small 
$\mu$ relative to $m_0$ can be maintained. The above result is a consequence of internal 
cancellation rather than fine tuning at the weak scale. Small $\mu$ has been utilized in 
several SUSY analyses of electroweak physics\cite{Feng:1999mn,Chattopadhyay:2007di,Baer:2012up,Baer:2013gva,Baer:2012cf}. We note here another phenomena which
allows for the generation of a natural  mass hierarchy where the squarks become massive with masses 
in the several TeV region while sleptons remain relatively light with masses in hundreds of GeV. This comes about for the case when the renormalization 
group evolution is driven by a heavy gluino\cite{Akula:2013ioa}, while $m_0$ is small which drives the squark masses into the TeV region due to their color interactions while the sleptons masses remain in the few hundred GeV region. Again the little hierarchy between slepton masses and squark masses is natural as it arises as a consequence of RG evolution.\\
}

{
  It is to be noted that mSUGRA is obtained under the assumptions of a universal Kahler metric over the set of 
  scalar fields in the visible sector and also a universal gauge kinetic energy function. However, the formalism
  of supergravity grand unification as given in Eq.(\ref{3}) and
  Eq.(\ref{1}) is general and non-universalities can be introduced 
  easily as noted in many subsequent works\cite{soniweldon,Matalliotakis:1994ft,Dimopoulos:1995mi,
  Nath:1997qm,Baer:2000cb,Corsetti:2000yq,Birkedal-Hansen:2002gkr}.  
  Further, it is to be noted that even though the scalar masses may be universal at the GUT scale they would not be universal at the electroweak scale. Thus, another issue of relevance for SUSY models relates to flavor changing neutral currents (FCNC).
 This comes about because
the sfermon soft masses and trilinear couplings need not be aligned
with the fermion Yukawa matrices which could in general introduce significant FCNC. 
However, FCNC are strongly constrained by data such as on 
 $K^0$--$\bar{K}^0$ mixing and on $\mu \to e \gamma$.
One approach to solving the SUSY flavor problem is the decoupling solution
where  the first and the second generation sfermions are much heavier 
(i.e., of size several TeV) than for the third generation,  which leads to a mass suppression of FCNC\cite{Cohen:1996vb}. 
The Seiberg alignment mechanism accomplishes such an alignment 
 using strong dynamics\cite{Nir:1993mx}.
}

\section{Tests of SUGRA GUTs}

\subsection{Gauge and Yukawa coupling unification}
 The history of the gauge coupling unification can be traced to the
  paper of Georgi, Quinn and Weinberg\cite{Georgi:1974yf}.  
   Their analysis demonstrated that the gauge couplings for the 
   strong, weak, and electromagnetic interactions were not constants 
   but dependent on the energy scale. Thus, the gauge couplings could 
   be extrapolated from low energy to high energy which would allow
   the possible test of their unification at  high scales.
    In the standard model, extrapolation does bring the gauge coupling constants close to each other at scales of $10^{14}-10^{15}$ GeV.
    However, the meeting of all three at one scale is missed by a wide margin.
  Within MSSM the gauge coupling unification does provide greater promise\cite{Dimopoulos:1981yj}. 
  Thus after the precision measurement of the gauge coupling constants 
  at LEP in 1991\cite{Amaldi:1991}, a large number of works appeared thereafter\cite{Ellis:1991,Langacker:1991an,Anselmo:1991,Arnowitt:1992,
   Ross:1992,Carena:1993, Hall:1993gn,Dasgupta:1995} that tested SUSY and SUGRA models with various degrees of precision. 
    However, it has been pointed out that gravitational smearing 
   from the unknown Planck scale physics may aﬀect predictions of an effective supersymmetric grand unification below the Planck scale \cite{Hill:1983xh,Shafi:1983gz,Hall:1992kq,Dasgupta:1995js}.   
   The precision measurements of the gauge coupling constants also have 
   important implications for string theory.
  Thus,  gauge coupling 
  unification in strings is analyzed 
in\cite{Dienes:1995,Dienes:1996du,Nilles:1997,Mayr:1993},
 and in F-theory grand unification in \cite{Blumenhagen:2009}. \\

 {Another important} aspect of SUGRA GUT is the possibility of Yukawa coupling unification for the third generation of quarks. Typically, this can occur for the case of large $\tan\beta$\cite{Ananthanarayan:1991xp} which could be as large as 
  $\tan\beta\sim 50$ but it can also arise for much smaller $\tan\beta$ in some models\cite{Babu:2006rp}.


\subsection{Upper limit on the Higgs boson mass in SUGRA}
Higgs boson mass prediction: 
  SUGRA models with radiative breaking of the electroweak symmetry
 allow  $A_0/m_0$ in a broad range, i.e., in the range $(-5,5)$ at the
 GUT {scale~\cite{Chattopadhyay:2007di}.} 
 Renormalization group  evolution then permits the trilinear coupling
 at the electroweak scale, $A_t$, to lie in a similar range that leads to $A_t$ of size $\mathcal{O}({\rm TeV})$.
 An important aspect of this observation is that a sizable $A_t$ makes 
 a corresponding sizable contribution to the mass of the Higgs boson 
for given values of $\mu, \tan \beta,m_0$. 
{ We note in passing that the possibility of generating large $A_t$ is  special to 
SUGRA models. Thus in gauge mediated breaking $A_t$ is loop suppressed relative to scalar masses\cite{Giudice:1998bp}.
In anomaly mediated breaking the $A_t$-term is also loop suppressed and  is not a free 
parameter\cite{Randall:1998uk} as in SUGRA models.} 
Returning to SUGRA models
since the light Higgs at the 
  tree level has a mass below $M_Z$ in SUGRA unified models, the loop
  correction can lift the Higgs mass substantially above $
M_Z$\cite{Berger:1989hg,Ellis:1990nz,Haber:1990aw,Okada:1990vk,Barbieri:1990ja}. 
  The largest loop correction arises from the stops ($\tilde t_1, \tilde t_2$),the supersymmetric partners of the top quark, with important contributions 
also from the supersymmetric partners of the b-quark, i.e. ($\tilde b_1, \tilde b_2$). Thus,  for large 
$\tan\beta$, the loop contributions from $\tilde b_i$ ($i=1,2$)
can become comparable to those from $\tilde t_i$ (i=1,2).
Turning the argument around, a large loop correction would imply
    a heavy sparticle spectrum,  and specifically heavy squarks, while the gaugino sector is not constrained. Using the argument above since one requires large loop corrections to pull the Higgs boson mass above its tree value, it points to the existence of at least some of the scalar sparticle spectrum
    to be heavy. Indeed, the Higgs boson mass loop correction allows the 
    {squarks to be as heavy or heavier than $10$ TeV~\cite{Chan:1997bi,Akula:2011aa} }
    while allowing the gauginos to be light. 
 Despite the fact that the {squarks} can become heavy up to 10 TeV and beyond, the solution is still natural within the hyperbolic branch\cite{Chan:1997bi} of radiative breaking. Of course, how one defines naturalness in this setting is somewhat subjective as there are various definitions of naturalness based on the specific criteria chosen, see e.g. the works\cite{Dimopoulos:1995mi,Feng:1999zg,Barbieri:2006dq,Baer:2013jla,Akula:2013ioa,Craig:2015pha,DelleRose:2017ukx}. Putting an upper limit on squark
 masses to be 10 TeV leads directly to an upper limit on the Higgs boson
 mass of $\sim 135$ GeV.
           Indeed, more rigorous analyses including constraints of electroweak
        symmetry predicted the Higgs boson mass to lie 
        below 130 GeV\cite{Akula:2011aa} (for related works see \cite{Baer:2011ab,Feng:2011aa,Heinemeyer:2011aa,Ellis:2012aa} and for early work on the Higgs boson 
        upper limit see \cite{Espinosa:1992hp,Casas:1994us}).    
      This result must be considered remarkable 
        in view of the fact that in the standard model the Higgs boson mass
        is constrained by unitarity and could be as high as $O(1)$TeV.
   The discussion above is  within the minimal SUGRA model.
   However, similar results hold in other  radiative breaking scenarios~\cite{Feldman:2011ud,Baer:2011sr}.
  We note in passing that  heavier Higgs boson  mass 
 can  also be obtained in  a variety of different ways, such as in
hierarchical breaking 
models~\cite{Tobe:2002zj,Arkani-Hamed:2004zhs,Kors:2004hz, Cabrera:2011bi}), by addition of vector like multiplets~\cite{Babu:2008ge}, in exceptional supersymmetric models\cite{Athron:2012sq}, among other possibilities.

\subsection{Vacuum stability in SUGRA vs SM}
For a Higgs boson mass of 125 GeV, vacuum stability in the standard model is not guaranteed within the current accuracy of measurement of the top 
quark mass. Thus, the Higgs potential in the SM is given by
$V(H)= -\mu^2  H^\dagger H + \lambda (H^\dagger H)^2$.  For $\lambda >0$,   a stable minimum is obtained on the electroweak scale with $v= 246$ GeV. 
{The RG evolution of the quartic coupling $\lambda$ is governed by the beta function 
$\beta_\lambda =d\lambda/dlog Q$,where $Q$ is the renormalization group scale.
 For the standard model Higgs potential it is given
by $\beta_\lambda \simeq (24 \lambda^2- 6y_t^4)/(4\pi)^2$
and is dominated by the top quark Yukawa coupling $y_t$.}
With the current experimental data on the Higgs boson mass and on the top quark mass, 
$\lambda(Q)$,  turns negative at high scale. This happens on a RG scale of $10^{10-11}$ GeV which leads to a new minimum which implies that the SM vacuum is 
in a metastable state. However, the tunneling rate is exponentially suppressed and leads to a lifetime for the SM vacuum, which is much larger than the age of the universe. For vacuum stability to be perfect, one would like $\lambda$ to remain positive up to the Planck scale, but this appears unlikely with the current measurements.  In supersymmetric models, the quartic couplings of the Higgs are determined by the electroweak gauge couplings  so that $\lambda_{\text{susy}}= (g^2 + g^{'2})\cos^2(2\beta)/4$ remains positive up to the Planck scale. Some of the relevant references can be found in \cite{Espinosa:2007qp,Gudrun}.
\subsection{Further tests of SUGRA}
There is a great amount of literature on SUSY and SUGRA, for which we refer to a  sample of books 
\cite{Nath:1983fp,Gates:1983nr,Mohapatra:1986uf,West:1990,Wess:1992cp,Weinberg:2000cr,Drees:2004jm,Baer:2006rs,Freedman:2012zz,Nath:2016qzm,Dreiner:2023yus} and reports~\cite{VanNieuwenhuizen:1981ae,Nilles:1983ge,Haber:1984rc,Martin:1997ns}.
In this section, we discuss the possible discovery of sparticles at the Large Hadron Collider (LHC), in precision experiments, and in astrophysical and cosmological data.
 The conventional signal of supersymmetry at the LHC is the production of  jets along with missing transverse energy. The current limits on gluinos and on squarks from jets + missing transverse energy are $\sim 2-3$ TeV in mass. 
 It is altogether possible that the gluinos and first generation squarks are heavy with masses as large as 10 TeV. However, it has been shown 
 that if one assumes a large gluino mass at the GUT scale while the rest of the mass parameters at the GUT scale are relatively much smaller, RG running will generate a split sparticle spectrum with heavy gluino and squarks, 
 while the electroweak gauginos and the sleptons remain light\cite{Akula:2013ioa}.
  In addition, the light spectrum could in general be compressed with some of
  the lightest sparticles clustering around the LSP. In this case if the
 mass difference between the next to lightest sparticle (NLSP) and the 
 lightest (LSP) is small, i.e., $\Delta m = m_{\text{sparticle}} - m_{\chi^0_1}$ is relatively small,  the missing transverse energy will be small and detection of susy 
  signal becomes more difficult.\\
  
   Furthermore, if $\Delta m$ is small enough so  that the NLSP is long lived, the decay of the NLSP could occur
  inside the {detector} with a displaced vertex and would be detected\cite{Aboubrahim:2019qpc}.
  Such a situation can occur for the case of a light chargino, a light stau
  or a light stop. Long lived squarks and gluinos can hadronize inside
  the chamber producing $\text{\bf R}$-hadrons.  
An explicit demonstration of the above scenario arises in gluino driven
   radiative breaking\cite{Akula:2013ioa} of the electroweak symmetry breaking where 
   {one} has a multi-TeV gluino and heavy
   first two generation squarks but light gauginos and sleptons in the few hundred GeV range and accessible 
   at the LHC. In such a scenario, the lighter stop could be light enough to be observable at the LHC or an NLSP which is long lived and decays inside the
   chamber producing $\text{\bf R}$-hadrons. Among other signatures, one of the first SUSY signatures discussed
was the trileptonic signal~\cite{Weinberg:1982tp,can3,Chamseddine:1983eg,can5,Dicus:1983cb}.
This signal arises when the light chargino decays into the second lightest neutralino, which then has a further decay into the lightest neutralino. Additional work on these signatures  was done
in ~\cite{Baer:1985at,bht}. These works involved the on-shell decays of the W-boson where $W\to \tilde \chi_1^{\pm} \tilde \chi_2^0$
followed by  on-shell decays of $\tilde \chi_1^{\pm}$ and  $\tilde \chi_2^0$. Subsequently, it was pointed out in ~\cite{Nath:sw} that the reach for  discovering a chargino can be significantly extended if one includes off-shell decays of the $W$ boson. These features
 are included in analyses of the trileptonic signal since then, see e.g., references 
~\cite{Baer:1994nr,Bornhauser:2011ab}. An in-depth discussion of the signatures of supersymmetry within the SUGRA models can be found in references \cite{SUGRAWorkingGroup:2000efm,Nath:2010zj}.\\ 

Indirect signatures of supersymmety can arise in precision experiments. Chief among these is the precision measurement of the muon anomaly $(g-2)_\mu$ which is ongoing and can provide a hint of supersymmetry if a significant deviation from the standard model prediction is found. This is so since supersymmetric electroweak contributions can be significant 
  and comparable to the standard model electroweak corrections\cite{Kosower:1983yw,Yuan:1984ww,Chattopadhyay:1995ae}.
  Currently, light sleptons (200-500 GeV)
   and light electroweak gauginos in a mass range similar to
   the one for sleptons,  are consistent with the  current experimental limits and can produce a detectable signal\cite{Aboubrahim:2021xfi} in future. 
   Another precision experiment of relevance
   is the experiment on the electric dipole moment (EDM) of the electron. EDMs for elementary particles in 
   SUSY can arise since SUSY has typically large CP violating phases via soft terms. 
   The ACME experiment\cite{Panda:2019dnf} that gives  
   $d_e<1.1\times 10^{-29}$ e·cm
severely constrains the SUSY CP-violating phases but allows
 TeV size superpartners with significantly less constraints on SUSY CP phases. An analysis within mSUGRA for a heavy sparticle spectrum is given in \cite{Abel:2005er}. Large CP phases can be managed within the cancelation mechanism\cite{Ibrahim:1997gj,Accomando:1999uj}.
   In the  presence of explicit CP violation, there is mixing of the CP even-CP odd Higgs bosons at the loop level which has interesting phenomenological implications 
\cite{Pilaftsis:1999qt,Pilaftsis:1998pe,Choi:2000wz,Ibrahim:2000qj}. It is possible that string models and string inspired supergravity models could generate a small CP phase, see, e.g., ref.\cite{Choi:1993yd}.\\

  We next discuss possible hints of SUGRA/strings in the decay of the proton.
The standard model has an exact $U(1)_B\times U(1)_L$ symmetry and proton decay is forbidden for renormalizable interactions. However, the electroweak sector of the SM contains nonperturbative configurations, i.e., instantons and sphalerons which violate $B$ and $L$ by 3 units so that $\Delta B=\Delta L=3$. They do not lead to proton decay.
For proton decay, one needs GUT models\cite{Weinberg:1981wj,Ellis:1981tv,Dimopoulos:1981dw,
 Chadha:1984cu,Nath:1985ub,Nath:2006ut}.
 However, non-SUSY $SU(5)$ is ruled out by the Super-K experiment
 \cite{Super-Kamiokande:2020wjk}, which puts a lower limit on the $e^+\pi^0$ mode so that 
   $\tau\big(p \rightarrow e^+ \pi^0\big) > 2.4 \times 10^{34}$ years~(90\% CL) 
   which is much higher than the one predicted by the non-SUSY SU(5),   
   while SUSY $SU(5)$ is consistent with the Super-K limit.  Similar  statements hold for $SO(10)$.
The distinctive proton decay mode that is a signature of SUSY/SUGRA is
$p\to \bar \nu K^+$ which is the discovery mode for SUSY GUTs\cite{Dimopoulos:1981dw}.
For this mode, the most recent  lower bound on the partial lifetime of Super-K\cite{Super-Kamiokande:2014otb} is
$\tau/B(p \to \bar{\nu} K^{+}) > 5.9 \times 10^{33}\ \text{years}~ (90\% \text{ C.L.})$ which is consistent with a variety of SUSY/SUGRA grand unified models\cite{Dev:2022jbf}. Proton decay has also been discussed 
in strings\cite{Friedmann:2002ty} and in \cite{Cvetic:2006iz} for intersecting D-brane models. It has been pointed out in the work of  Friedmann and Witten that in certain M-theory compactifications
$p\to e^+_R+ \pi^0$ is suppressed relative to 
$p\to e^+_L+ \pi^0$. Thus, the observation of the  proton decay mode $p \to \bar{\nu} K^{+}$ would be an indication of SUSY/SUGRA-GUTS and the observation of $\to e^+ \pi^0$ with the dominance of $p\to e^+_L+ \pi^0$
 would be a strong hint of its stringy origin. 
We note that if R-parity is violated, such as can occur in spontaneous breaking, the signatures of supersymmetry will be drastically 
affected\cite{Barger:2008wn,Feldman:2011ms}.

\section{SUGRA in Cosmology}
SUGRA provides a UV-motivated framework for the study of 
inflation, dark matter, baryogenesis, moduli stabilization, and reheating. 
Here, {the structure} of the Kahler  potential, superpotential, gauge kinetic function, and SUSY-breaking determine cosmological phenomena. We discuss some specific examples below.\\
 {\it Inflation:}
  Big Bang cosmology, though highly successful in explaining a vast amount of data, is not a complete theory, specifically as it relates to the very
early history of the universe.  Big Bang cosmology is known to have at least three puzzles. These include the flatness problem,  the horizon problem, and the monopole problem. 
 Inflation solves these problems under the condition that $H=\dot a /a$, 
 where $a$ is the scale factor in the Friedmann-Robertson-Walker line 
 element,  is a constant at some period in the early history of the universe.
 In that case $a(t)\sim a(t_0) e^{Ht}$. In this case there would be 
 an exponential increase in the scale factor, which helps resolve the
  Big Bang cosmology problems.  In the context of field theory involving a scalar or a pseudo-scalar field $\phi$, such an exponential increase can be manufactured if the potential for $\phi$ satisfies 
  slow-roll conditions. Thus, define the 
 parameters $\epsilon, \eta, \xi^2$, where
 $\epsilon \equiv = ({V'}/{\sqrt 2\kappa V})^2$,~
 $\eta\equiv |{V''}/{\kappa^2 V}|$, 
 $\xi^2\equiv ({V'V''}/{\kappa^4 V^2})$, {and where $V'$ means a derivative of $V$
 with respect to the inflaton field},
 all of which are  required to be small for slow roll, i.e., $(\epsilon, \eta, \xi^2)<<1$.
 In supergravity, inflation can occur via either an
 $F$-term or a $D$-term. For the F-term inflation, one finds that 
 $V''=\kappa^2 K'' V+ \cdots$ which gives $\eta\sim 1$ and
 does not satisfy the slow role condition, i.e.,
 $\eta<<1$. This is the so called $\eta$ problem of supergravity inflation.
 One simple way to overcome this problem is to impose the Nambu-Goldstone shift symmetry  under which the Kahler  potential is invariant, i.e., one assumes  the Kahler  potential to be  of the form
 $K(\phi+\phi^*)$ \cite{Kawasaki:2000yn,Nath:2017ihp}. In this case the Kahler  potential is independent of
 the inflaton field if we assume the inflaton field to be the imaginary part of 
 $\phi$. 
 For D-term 
 inflation\cite{Halyo1996,Randall1997,Kors:2004hz}, one does not have an
 outside factor of $exp(\kappa^2 K)$ and therefore does not have the $\eta$ problem. In addition, consistent inflation has been discussed in no-scale supergravity  models\cite{Ellis:2013xoa}, 
  in DBI models \cite{Alishahiha:2004eh,Nath:2018xxe}  
  and in superconformal models\cite{Kallosh:2013lkr} which are a class of models where inflation arises after breaking of the superconformal symmetry which leads to supergravity like models.\\

{\it 
{Baryogenesis in SUGRA}:}  
 Baryogenesis involves three conditions enunciated by Andre Sakharov,
 which are the violation of baryon number, the violation of CP, and the departure from thermal equilibrium\cite{Sakharov:1967}.  
 Electroweak phase transitions in the standard model cannot 
 accomplish this, as there is not sufficient CP violation in SM, and the 
 phase transitions are second order. In MSSM electroweak baryogenesis
 is a possibility in a first-order electroweak phase transition but 
 there  are stringent constraints due to the Higgs boson mass
 and the experimental lower limit constraints on the light stop. Thus
  electroweak baryogenesis in MSSM though not ruled out appears less 
  likely as the preferred mechanism for baryogenesis. 
However, there are other avenues
for generating baryogenesis in a supersymmetric framework, one of which is the 
Affleck-Dine mechanism\cite{Affleck:1984fy}.
{The current favored mechanism for the baryon asymmetry is via leptogenesis where
lepton asymmetry is generated first and then transferred to baryon asymmetry via sphaleron interactions.
The initial proposal on baryogenesis via leptogenesis was made in reference\cite{FukugitaYanagida1986}, 
and extensions to SUSY case were given in several later works \cite{Luty1992},\cite{Dine:1995kz},
\cite{CoviRouletVissani1996}, \cite{BuchmullerPlumacher1996}, \cite{Grossman2003}.}
Further, baryon asymmetry can naturally arise in SUGRA
if the visible sector is coupled to the hidden sector.
Here, a heavy field in the hidden sector can decay 
 partially into a visible sector field that carries 
 a lepton number and into a hidden sector field that carries an equal and opposite quantum number.
 The lepton number in the visible sector is 
 then  converted into a baryon number through sphaleron effects\cite{Feng:2013wn,Feng:2013zda}. For a review of  baryogenesis, see \cite{Dine:2003ax}. 
 
 \subsection{SUGRA dark matter: gravitino, neutralino, axion}
   The gravitino was discussed as a possible candidate for dark matter by  Primack and Pagels\cite{Pagels:1981ke} who computed its relic density. Assuming that 
   {the gravitino} has a thermal production and under the constraint that the gravitino  relic density does  not overclose the universe, they deduced a bound on its mass so that $m_{3/2} \leq O(1 \text{keV})$. More recently,
 using cosmological constraints to avoid suppression of small-scale 
  structure, Sato et al.\cite{Osato:2016ixc}  obtained a much stricter bound on the gravitino mass of $m_{3/2} \leq 4.7 \text{eV}$  assuming standard cosmology and thermal production. This bound is 
   orders of magnitude smaller than the one obtained by Primack and Pagels. mSUGRA offers another candidate for dark matter, i.e., the lowest mass supersymmetric particle (LSP) if it is charge neutral.
  Thus in mSUGRA under R parity conservation (for a review see \cite{Mohapatra:2015fua}) if the LSP is a neutralino
  it would be perfectly stable and a candidate for dark matter.
  Further, in mSUGRA  under the evolution of RG from the GUT scale down to the electroweak scale, the lightest neutralino turns out to be the LSP\cite{Arnowitt:1992aq}. Thus, soon after the formulation of 
  SUGRA models, Goldberg\cite{Goldberg:1983nd} proposed the neutralino as a candidate for dark matter. This proposal was followed by numerous authors
  and has yielded a rich phenomenology both in the astrophysical searches for dark matter and for the discovery of dark matter at colliders, which would manifest as missing energy. For further work see, e.g. 
\cite{Krauss1983Ino,Ellis:1983ew,Nath:1992ty,Baer:2004fu},
  and for a comprehensive review of supersymmetric dark matter, see \cite{Jungman:1995df}.
 Here, one finds that RG evolution of matter and gauge particles yields the
 lightest neutralino as the LSP whose composition is model dependent as it can be close to being a Bino, a Wino or a higgsino or a linear combination thereof. 
 It is also of interest to note that within mSUGRA, that measurements of neutralino production at the LHC allow one to make predictions on the amount of dark matter in the universe\cite{Arnowitt:2008bz}.\\ 


More recently, the hidden sector has come to play a role in dark matter analyses. Thus,
conventional dark matter analyses depend on the freeze-out mechanism\cite{Lee:1977ua} for the
generation of dark matter. Here, the weakly interacting particles are in thermal
equilibrium until they fall out of equilibrium at the freezeout temperature, which
contributes to the freezeout relic density. 
For feebly interacting dark particles, an
 initial density of them does not exist because  of their feeble interactions. Rather, they are produced by the freeze-in mechanism\cite{Hall:2009bx,Belanger:2018mqt,Tsao:2017vtn}
 arising from the decay of the next-to-lightest supersymmetric particle which 
generates the relic density. More realistically, the relic density may be a combination of 
the freeze-in and freeze-out mechanisms. Thus, suppose that the LSP resides in the hidden sector
and has feeble interactions with the visible sector. Next, suppose that the {NLSP} is an
MSSM particle that decays into the LSP.  
In this case, the freeze-in contribution would arise from the decay of the NLSP. However, {the} NLSP which would
be in thermal equilibrium initially, but finally would fall out of equilibrium at the freeze-out temperature and then eventually decay into the hidden sector LSP's. A specific example of this phenomenon is the decay of an NLSP stop into
 top and the hidden sector LSP, i.e.
 $\tilde t \to  t+ \tilde\xi^0$, where $\tilde\xi^0$ is the dark matter particle in the hidden sector.  In this case, the relic density of the LSP will
consist of the sum of the freezein component 
and a freezeout component\cite{Aboubrahim:2019kpb}. 
 There are several experiments investigating neutralino dark matter: {\scriptsize LUX-ZEPLIN (LZ), XENONnT, PandaX-4T}. The mass limits are model dependent. The neutralino could be  of low or high mass in the range of $O(1)$ GeV to O(1) TeV mass. 
 {The latest results from LZ\cite{LZ:2024zvo} are the most stringent on WIMP-nucleon cross section 
 and severely constrain the MSSM parameter space, particularly with significant 
 Bino-Higgsino mixing. The detection sensitivity of the experiment is approaching the 
 so-called neutrino floor where $\sigma_{SI}\sim 10^{-49}$cm$^2$ and where neutrino  scattering 
 becomes an irreducible background.}
 Aside from the gravitino and the neutralino, an axion or axino could be a dark matter candidate. Specifically, the axion could be ultralight (fuzzy dark matter), with mass as low as $10^{-21}$ eV. Such a mass arises naturally in particle physics models, including those from strings and from  SUGRA models\cite{Kim:2015yna,Hui:2016ltb,Halverson:2017deq,Aboubrahim:2024spa}.
\subsection{Dark energy in SUGRA}
The standard cosmological model accommodates dark energy through
the cosmological constant $\Lambda$, 
which is related to the vacuum energy $\rho_{\text{v}}$ via the relation
   $\rho_{\text{v}}= \Lambda M_{\text{Pl}}^2/ 8\pi$. The vacuum energy is 
   very small, i e., $\rho_{\text{v}}  \sim (2.3 meV)^4$.
  The smallness of the vacuum energy presents a puzzle, and there are a 
  large number of  scholarly works on the possible origin of its smallness (for a review, see\cite{Weinberg1989}). In the standard model, vacuum energy is huge i.e., $O(M^4_{\text{Pl}})$. In SUSY/SUGRA there is some relief in that 
  the vacuum energy is $O(m^2_{3/2}M^2_{\text{Pl}})$ which is still very far from the observed value (for analyses in SUSY/SUGRA and strings see 
  references \cite{Nicolai1982,Bagger1994,Pilo2000,Douglas2007,Witten2001}).
 {One proposed explanation, primarily advocated by Weinberg\cite{Weinberg1989}, is  the anthropic principle, based on the following idea: Suppose there exists a multiverse where the cosmological constant takes different values in different regions. If the cosmological constant were much larger and positive, the universe would expand too rapidly for galaxies to form, and if
   it were large and negative, the universe would rapidly recollapse before structure formation.
     This argument leads only to a narrow range of values which allow for the formation of galaxies, stars, and ultimately life.}
  However, it is entirely possible that dark energy is not a constant but a dynamical field, called quintessence, 
  as proposed by Peebles and Ratra\cite{RatraPeebles1988,PeeblesRatra2003} and there is much further work on this idea, see, e.g., \cite{Caldwell1998,Wang1999}. The incorporation of 
  quintessence in SUGRA and strings has been discussed by 
  several authors \cite{ChiangMurayama2018,ValeixoParameswaran2020,  
PandaSumitomoTrivedi2010}. Recent analyses by 
 DESI DR2\cite{DESI:2025zgx}, show that dark energy may be time dependent,
 giving support to the idea of a dynamical origin of dark energy\cite{Aboubrahim:2025usl}. For reviews related to dark energy 
 and intertwining of particle physics and cosmology, see \cite{Abdalla:2022yfr,CosmoVerseNetwork:2025alb}. 

\section{SUGRA and strings} 
\subsection{Variety of supergravities}
The {connection} between superstring and SUGRA was noticed early
on in the work of Candelas et al., \cite{candelas:1985en}. We discuss this connection in further detail below. Thus, for each of the string theories, Type IIA, Type IIB, Type I, and heterotic strings, one has associated
supergravity theories in the low energy limit. Vanishing of the Weyl anomaly fixes the critical spacetime dimension to be ten for superstrings \cite{Polchinski1998a,GSW1987} and consequently for the associated
supergravities in the low energy limit. The mass spectrum of strings consists of  massless states and an infinite tower of massive states with masses that are typically $m^2 \sim {N}/{\alpha'}$, where $\alpha'$ is the inverse string tension.
In the low energy limit when  $E \ll {1}/{\sqrt{\alpha'}}$, only the massless modes survive and the massless spectrum matches the massless spectrum of the corresponding 10 dimensional supergravity.
There is a common Neveu-Schwarz-Neveu-Schwarz (NS-NS) sector for all superstring theories.
The massless fields of this sector consist of $g_{\mu\nu}, B_{\mu\nu}, \Phi$, where $g_{\mu\nu}$ is the spacetime metric, $B_{\mu\nu}$ is the antisymmetric two-form, and $\Phi$ is the dilaton.
The above fields constitute  the gravitational sector of a ten-dimensional supergravity theory. 
The requirement of conformal invariance of the 
worldsheet leads to spacetime equations of motion derivable from a supergravity action, and there is suppression by powers of $\alpha'$ of higher-derivative contributions \cite{Callan1985,Fradkin1985}. The Ramond-Ramond (R-R) sectors of the various superstring theories are different. Thus, the R-R sector of Type II A string, which is a theory of closed oriented strings with non-chiral supersymmetry, consists of a one-form $C_1$ and a three-form $C_3$.
The Type IIA supergravity is  non-chiral, has $N=2$ supersymmetry with 32  supercharges\cite{Campbell1984} and arises from a dimensional reduction of 11-dimensional supergravity on a circle\cite{Witten1995}.\\

Type IIB string theory is a chiral theory of closed, oriented strings with $\mathcal{N}=2$ supersymmetry. Its massless R-R sector consists of a scalar $C_0$, a two-form $C_2$, and a self-dual four-form $C_4$. Its low energy limit is Type IIB supergravity \cite{Schwarz1983}, is free of anomalies due to cancelations.
The Type I string has open and closed unoriented strings.  Its massless spectrum in the closed string sector includes  a supergravity multiplet, and the massless spectrum in the open-string sector includes gauge fields.
The Green--Schwarz mechanism for anomaly cancelation constrains the gauge group to $SO(32)$ or $E_8\times E_8$ \cite{GreenSchwarz1984}. For Type I string the gauge group is limited to $SO(32)$.
  Its low energy limit is $\mathcal{N}=1$ ten-dimensional supergravity coupled to $SO(32)$ 
  superYang-Mills. Heterotic string theories are purely closed-string theories involving left-moving bosonic degrees of freedom and right-moving supersymmetric degrees of freedom. The anomaly cancelation allows for the  gauge groups $SO(32)$, and  $E_8 \times E_8$.
The low energy limit of this theory is  $\mathcal{N}=1$ ten-dimensional supergravity coupled to superYang--Mills \cite{Gross1985,Hull1985}. 
 One of the interesting results in the Green and Schwarz analysis \cite{GreenSchwarz1983} is that the  superstring theory in the zero slope limit gives the ten-dimensional supergravity
 \cite{Chamseddine1981}
 coupled to the supersymmetric Yang-Mills theory with the presence of a higher-derivative term
proportional to the Lorentz-group Chern-Simons form, i.e.,

$$\partial_{[P} B_{MN]} \to \partial_{[P} B_{MN]}+ \omega_{PMN}^{(Y)}- 
{\omega_{PMN}^{(L)}},$$ 
where $\omega_{PMN}^{(Y)}$ is the Yang-Mills and $\omega_{PMN}^{(L)}$ is the Lorentz-group Chern-Simons form. Previous works \cite{Chamseddine:1981jr,Bergshoeff:1981um,Chapline:1982ww} which preceded the 
Green and Schwarz work, involved only the Yang-Mills Chern-Simons term.
Inclusion of the Lorentz-group Chern-Simons form implies that the low energy supergravity Lagrangian need to be computed up to order 
$\kappa^2$ whereas the previous formulations were complete only up to 
order $\kappa$. Thus, after the Green-Schwarz work, 
the computation of the 10D Sugra Lagrangian to  order $\kappa^2$ was accomplished
in several works \cite{Chamseddine:1986gj,RomansWarner1986,GatesNishino1986,BergshoeffSalamSezgin1987}.


\subsection{Green-Schwarz anomaly,  Stueckelberg extensions of SUGRA, and milli charges}
The Stueckelberg mechanism is a phenomenon in which an Abelian gauge boson can become massive by absorbing an axionic field without spontaneous symmetry breaking\cite{Stueckelberg:1938zz,Ogievetskij:1962mba} as happens for the Higgs boson \cite{Higgs:1966ev,Higgs:1964pj,Higgs:1964ia,Englert:1964et, Guralnik:1964eu}. 
Specifically, this occurs if an abelian gauge field $A_\mu$ is replaced by  $A_\mu + \frac{1}{m}\partial_\mu \sigma$. Here, $\sigma$ is the axionic field and generates the longitudinal mode for $A_\mu$. Furthermore, under the
gauge transformation $\delta A_\mu=\partial_\mu \lambda, ~~\delta \sigma= -m \lambda$, the combination $A_\mu + \frac{1}{m}\partial_\mu \sigma$
remains invariant. The mass parameter $m$ that appears along with the 
axionic field was identified as topological\cite{Allen:1990gb}.
 In addition, a Lagrangian built from a massive abelian vector field using the Stueckelberg mechanism has been shown to be renormalizable\cite{Slavnov:1972qb,Veltman:1968ki}.
Stueckelberg couplings arise in the reduction of  higher dimensional
  supergravity theory and in string theory. Thus, for example, consider the 
  reduction of 10 dimensional supergravity coupled to supersymmetric 
  Yang-Mills fields in the presence of internal gauge fluxes. Here
  the 10-d kinetic term for the tensor $B_{IJ}$, which is anti-symmetric, 
  couples to the Yang-Mills Chern-Simons term, which is of the form
  $\partial_{[I} B_{JK]} + A_{[I} F_{JK]} + \cdots$.  The use of the expectation value of the
internal gauge field strengths $<F_{ij}>\neq 0$ in dimensional reduction leads to the form $\partial_\mu B_{ij} + A_\mu F_{ij}$. If we 
  identify $B_{ij}$ as the axionic scalar and $<F_{ij}>$ as the mass parameter
 $m$, we arrive at the combination $\partial_\mu \sigma + m A_\mu$.
  In the Lagrangian, this combination leads to a coupling $mA_\mu \partial^\mu \sigma$. This coupling also enters the Green-Schwarz anomaly cancelation. In dimensional reduction, the parameter is typically of the compactification scale in some cases\cite{Ghilencea:2002da}, but in principle it could be different\cite{Ibanez:1998qp}.\\

{Stueckelberg extensions} of the standard model and of MSSM 
have been obtained using the mechanism discussed above with the phenomenon of mixing of the hypercharge with an additional 
     $U(1)_X$ gauge field which could arise from a hidden sector\cite{Kors:2004dx,Kors:2004ri}. Kinetic mixing of two $U(1)$ gauge fields has been discussed in previous work in the field
     theory context in \cite{Holdom:1985ag} and in the string context in \cite{Dienes:1996zr}. However, in Stueckelberg extensions one has mass mixings between the gauge fields.
     Thus, one may consider the new gauge group 
$SU(3)\times {SU(2)_L\times U(1)_Y}\times U(1)_X$.
   Mixings between $U(1)_Y$ and $U(1)_X$  lead to a mass term of the form
   $- \frac{1}{2} (\partial_{\mu}\sigma  + M_1 A_{\mu} +M_2 B_{\mu})^2$.
 Including $SU(2)_L\times U(1)_Y$ electroweak breaking gives a 
 $3\times 3$ dimensional mass matrix for the gauge bosons 
 $A^3_{\mu}, B^\mu, A^\mu$.  
In addition, the Stueckelberg extended model predicts a sharp  $Z'$ resonance, which is a signature of this class of extended models.
 The prediction of a sharp resonance is in contrast to other $U(1)$-  extensions where $Z'$ has a high mass, i.e., ${M_{Z}}/{M_{Z'}} <<1$ 
 needed to conform to the stringent electroweak constraints on a new $Z'$ 
boson\cite{Cvetic:1996mf,Cvetic:1995rj,Faraggi:1996kk,Barger:2003hg, Barger:2003hg,Anoka:2004vf,Gherghetta:1996yr,Barger:1989sf,Hewett:1988xc}. In contrast, Stueckelberg $Z'$ can be low lying, 
   consistent with electroweak constraints, since 
   $\delta \equiv {M_2}/{M_1}$ can be chosen to be small.
   The Stueckelberg model can be tested at FCC-ee\cite{Arduini:2025hor}, which can detect a sharp $Z'$ resonance predicted by the model.\\

 A further phenomenon associated with Stueckelberg extension relates to the generation of milli-charges
  which are forbidden in the standard model.
  They can arise in the kinetic  mixing of two massless gauge fields\cite{Holdom:1985ag} which, however,
  is inconsistent with observation and is thus ruled out. However, milli charges may arise in quantum gravity and strings.  
      Specifically, Stueckelberg extensions of the standard model and of SUGRA  
 predict milli-electrically charged particles if the particles carry a $U(1)_X$ charge\cite{Kors:2004dx,Kors:2004ri,Cheung:2007ut,Feldman:2007wj}.
 In this case, the Stueckelberg mass mixing mechanism leads to milli-charged particles in the hidden sector. We note that one of the favored solutions to the  
 EDGES anomaly\footnote{
 {\scriptsize EDGES (Experiment to Detect the Global Epoch of Reionization Signature) is a radio astronomy experiment designed to measure signals from roughly a few hundred million years after the Big Bang. EDGES found that the baryons in the early universe appeared to be colder than predicted which is known as the EDGES anomaly.
 A small amount of milli-charged dark matter from the hidden sector cools the baryon via Rutherford scattering.}} is the presence of a small amount of milli-charged 
 dark matter\cite{Creque-Sarbinowski:2019mcm,Aboubrahim:2021ohe} that
 can cool the baryons in the early universe, consistent with EDGE's 
  observation\cite{Bowman:2018yin}.
Currently, there are several experiments searching for milli-charged 
particles\cite{Prinz:1998SLAC,Haas:2015MilliQan,Ball:2020MilliQan,Bailloeul:2025fde,Gninenko:2025aek}.


\section{Future prospects}   
SUGRA models allow for a description of physics all the way from the 
electroweak scale up to the grand unification scale. Currently, there is no alternative formalism that can do that. As noted above SUSY and SUGRA arise as a low energy limit of strings below the Planck scale. This feature is of import in view of the prevalent view  
 that not all consistent theories of particle physics are remnants of quantum gravity, and most of them belong to the so called swampland~\cite{ooguri_vafa_2007}. Acceptable particle physics theories, which may be considered as quantum gravity remnants, belong to what is known as the landscape. 
 {The landscape idea takes on a concrete form in the context of string theory which admits a vast number of possible vacuum solutions as many as $10^{500}$ or more. 
  Each vacuum corresponds to a different way of compactifying the extra spatial dimensions and choosing background fluxes, and each vacuum gives rise to different low-energy physics.
  Our universe is just one vacuum state in the vast number of vacuum states of the string landscape.}
  Since SUGRA is expected to arise as a low energy limit of a superstring, it elevates SUGRA as the right candidate for an effective theory of particle physics, which would 
be a quantum gravity remnant.
 The next challenge is, of course, to find the exact
string embedding of the m(n)SUGRA models. There is another aspect of SUGRA 
that supports this scenario. This relates to the hidden sector 
 needed for the gravity mediated breaking of supersymmetry in SUGRA.
 Hidden sectors also exist in strings, such as the heterotic string $E_8\times E_8$, where one of the $E_8$ is identified with the hidden sector. 
 Further, over the past decades, since the emergence of SUSY/SUGRA, one has seen an enhanced intertwining of particle physics and cosmology
 which is manifest in the analyses of inflation, dark matter, and dark energy pointing to the possibility of their  common origin. 
  As discussed in the body of the paper, there is already a  significant indirect support for SUSY/SUGRA, including solution to the gauge hierarchy problem, unification of gauge couplings, {radiative breaking of the electroweak symmetry driven by the heavy top quark,}
     prediction of the upper limit of the Higgs boson mass, vacuum stability of the universe, and consistency with the proton lifetime limits. Currently, the major missing piece for the validation of  SUSY/SUGRA is experimental observation of sparticles.
 Here, our hope is pinned on the high luminosity LHC RUN 4
 which is projected to begin its operation in mid-2030 and is expected to accumulate 3000-4000 fb$^{-1}$ of integrated luminosity.\\


Acknowledgments: PN research was supported in part by the NSF Grant PHY-2209903.


\section{Appendix A: The contraction 
of Gauge Supersymmetry to 
of Standard Supergravity}

The idea that led to the development of gauge supersymmetry in 1975 \cite{Nath:1975nj} was based on the premise that fundamental theories of physics
in the realm of energy scales we deal with should be gauge theories.
This is the case for all successful fundamental theories  of physics, i.e., Maxwell-Dirac, Einstein gravity theory, the standard model of particle physics, all of which are gauge theories. However, supersymmetry until 1975 was a global symmetry and if it were truly a fundamental symmetry of nature, it should be gauged. This was the motivation for the work of \cite{Nath:1975nj}. The procedure used to achieve this was to extend Einstein gravity theory in ordinary space to gauge supersymmetry in superspace and create a geometry of curved superspace~\cite{Nath:1975nj}.
  This formulation used a single tensor superfield $g_{\Lambda\Pi}(z)$
in superspace consisting of bose and fermi co-ordinates, i.e.,
$z^\Lambda=(x^\mu, \theta^{\alpha i})$,  where $x^\mu$ are the bose co-ordinates of ordinary space-time
and $\theta^{\alpha i}$ are anti-commuting Majorana fermi co-ordinates \footnote{    
The global and local space labeling conventions used here are as follows: Greek letters label global
coordinates and latin letters the tangent space. Capital letters run over the full Bose and Fermi indices of superspace. Early indices, e.g., $\alpha,\beta,\cdots$  or $a,b,\cdots$ are used for Fermi coordinates and  the late indices such as $\mu,\nu,\cdots$ or $m,n,\cdots$ are used for Bose coordinates.
The global superspace coordinate is $z^A=(x^\mu, \theta^{\alpha i})$, where $\alpha=1,2,3,4$ is the 
Fermi Majorana index and $i=1,\dots,N$ is the internal symmetry index. The sign factor $(-1)^A, A=(0,1)$ where $0$ for bose coordinates and $1$ for Fermi coordinates.
 The Lorentz metric $\eta_{mn}$ has sigmature $+2$ while the matrix $\eta_{ab}$ is 
 given by the charge conjugation matrix $C$ so that $\eta_{ab}= (-C^{-1})_{ab}$.}.
The analysis of \cite{Nath:1975nj} is based on invariance of the line element 
$ds^2=dz^\Lambda g_{\Lambda\Pi}(z) dz^\Pi$ under the transformations $z^{\Lambda}= z^{'\Lambda} + \xi^\Lambda(z)$ which lead to the dynamical equations 
\begin{align}
R_{AB}=\lambda g_{AB}.
\end{align}
where $R_{AB}\equiv (-1)^c R^C_{~~ABC}$
 where $g_{AB}=(g_{\mu\nu},g_{\mu\alpha}, g_{\alpha\beta})$ and correspondingly $R_{AB}= (R_{\mu\nu},
 R_{\mu\alpha}, R_{\alpha\beta})$. The equations of motion can be derived from an action integral given by\cite{Arnowitt:1975xg}
 \begin{align}
  A = \int d^8 z \sqrt{-g} (R-2\lambda), 
 \end{align}  
where $R= (-1)^a g^{AB} R_{BA}$, and  $\sqrt{-g}$ is given by\cite{Arnowitt:1975xg}

\begin{align}
\sqrt{-g}= [-(det g_{\mu\nu}) (det(g^{\alpha\beta})]^{1/2}.
\end{align}
Defining $R^{AB}= g^{AC} R_{CD} g^{DB}$ one can deduce the following Bianchi identity
\begin{align}
 (-1)^b [R^{BA}- \frac{1}{2} g^{BA} R]_{;B}=0.
\end{align}
which constitute the general  superspace co-ordinate   transformations and
induce  transformations of the metric $g_{\Lambda\Pi}(z)$ of the form 
\begin{equation}
\delta g_{\Lambda\Pi}(z)= g_{\Lambda\Sigma} \xi^\Sigma_{,\Pi} + (-1)^{\Lambda+ \Lambda\Sigma} \xi^\Sigma_{,\Lambda}
\xi_{\Sigma\Pi} + g_{\Lambda\Pi,\Sigma} \xi^\Sigma\,. 
\label{deltagAB}
\end{equation}
where $(-1)^{\Lambda}= 1$ when $\Lambda$ is for the bosonic case  and $(-1)^{\Lambda}=-1$ when $\Lambda$ is for the fermionic case.
The tangent space group of this theory is $OSp(3,1|4N)$ with the tangent space metric given by

\begin{equation}
\eta_{AB} =\text{diag}\left(\eta_{mn}, k\eta_{ab}\right).
\label{gb1}
\end{equation}
Here $\eta_{mn}$ is the Bose space metric and $\eta_{ab}$ is 
the fermi space metric defined by $\eta_{ab}= -(C^{-1})_{ab}$ where $C$ is the charge conjugation matrix. We note that the fermionic space brings in a new parameter $k$ which is arbitrary .
Using this tangent space metric one defines the super vierbein  by
\begin{equation}
g_{\Lambda \Pi}(z) = V^A_{~~\Lambda}(z) \eta_{AB} (-1)^{(1+B) \Pi} V^B_{~~\Pi} (z)
\end{equation}
 Transformations of global supersymmetry correspond to  $\xi^\Lambda=(\xi^\mu, \xi^\alpha)$ 
 where
\begin{equation}
\xi^\mu =i \bar\lambda  \gamma^\mu \theta, ~~~ \xi^\alpha= \lambda^\alpha\,.
\label{xi-1}
\end{equation} 
Here $\lambda^\alpha$ are infinitesimal anti-commuting  parameters with no $x$ dependence. 
 For the global supersymmetry transformations of Eq.(\ref{xi-1}), the invariant line element is given by the metric components
\begin{align}
g_{\mu\nu} &= \eta_{\mu\nu}\,, \nonumber\\
g_{\mu\alpha} &= -i(\bar\theta \gamma^m)_{\alpha} \eta_{m \mu}\,,\nonumber\\
g_{\alpha \beta}&= k \eta_{\alpha\beta} + (\bar \theta \gamma^m)_{\alpha} (\bar \theta\gamma_m)_\beta\,.
\label{global}
\end{align}\\
 To include gravitation, we introduce the vierbein field $e^\mu_a$
 and the metric is defined so that $g_{\mu\nu}(x) = e_{a\mu} e^a_{\nu}$.
 These fields must enter the metric in a way that Einstein gauge transformations $\xi^\mu(z)=(\xi^\mu(x),\xi^\alpha=0)$ and the local vierbein rotations $\xi^\mu(z)=0, \xi^\alpha(z)=\frac{i}{4}
 \omega_{ab}(x) (\sigma^{ab} \theta)^\alpha$ are produced by Eq. (\ref{deltagAB}). The constraints above
 with respect to these subgroups are called `gauge completeness' in~\cite{Nath:1976ci,Arnowitt:1978jq,Nath:1979yn}. 
 Thus, the metric which is gauge complete with respect to Einstein and local vierbein transformations is given by 
 \begin{align}
 g_{\mu\nu}(z)&= g_{\mu\nu}(x) - i \bar\theta \gamma_{(\mu}\Gamma^{(0)}_{\nu)} \theta 
 -k \bar \theta \Gamma_\mu^{(0)} \Gamma_{\nu}^{(0)} \theta + \cdots\,, \nonumber\\
 g_{\mu\alpha}(z) &= -i (\bar \theta \gamma^a)_{\alpha} e_{a\mu} (x) - k (\bar \theta 
 \Gamma^{(0)}_\mu)_\alpha + \cdots 
 \label{gcomp}
 \end{align}
 where 
 $\Gamma_\mu^{(0)}= \frac{1}{4} \omega^{(0)}_{ab\mu}\sigma^{ab}$ where $\omega^{(0)}_{ab\mu}$ is defined so that 
 \begin{align}
 \omega^{(0)}_{abc}\equiv \frac{1}{2} \left[\Omega_{cab} + \Omega_{bca}-\Omega_{abc}\right]\,
 \end{align}
 where $\Omega_{abc}= e_a^{~~\lambda} e_{b}^{~~\rho} (e_{c\rho, \lambda} - e_{c\lambda,\rho})$ and $\omega^{(0)}_{ab\mu}=  \omega^{(0)}_{abc}e^c_{~~\mu}$.
 
  However, the metric of Eq.(\ref{gcomp}) is not yet local supersymmetry covariant. For that we need to 
 include a spin $3/2$ Majorana field and make $\lambda$ a function of $x$. Thus the local supersymmetry transformations require
 \begin{align}
 \tilde{\xi}^\mu &=i\bar \lambda(x) \gamma^\mu \theta + \frac{1}{2} (\bar \psi_a \gamma^\mu \theta) 
 (\bar \lambda(x) \gamma^a \theta),\nonumber\\
 \tilde {\xi}^\alpha&= \lambda^{\alpha}(x) - \frac{i}{2} \psi_a^{~~\alpha} (\bar \lambda(x) \gamma^a \theta),
 \label{no-k}
 \end{align}
 where $\psi_a= \psi_\mu  e_a^{~~\mu}$.  From Eq.(\ref{no-k}) we can compute the components of the 
 metric that are independent of $k$ that are
 \begin{align}
 \tilde{g}_{\mu\nu}(z)&= g_{\mu\nu} (x) + i\bar{\psi}_{(\mu}\gamma_{\nu)} \theta - i \bar\theta \gamma_{(\mu}
 \Gamma_{\nu)}\theta\nonumber\\
  & ~~~-\frac{1}{2}   (\bar \theta_{(\mu}\gamma^a \theta)  (\bar \theta_{\nu)}\gamma_a \theta)\,,  \nonumber\\
 \tilde {g}_{\mu\alpha}& = -i (\bar \theta \gamma^a)_\alpha e_{a\mu} + (\bar \psi_\mu \gamma_a  \theta) 
 (\bar \theta \gamma^a)_\alpha\,, 
 \end{align}
 and $\tilde{g}_{\alpha\beta}$ is as given in Eq.(\ref{global}). In the above $\Gamma_\mu$ is given by 
 \begin{align}
 \Gamma_\mu&= \frac{i}{4} \sigma^{ab} \omega_{ab\mu},\nonumber\\
\omega_{ab\mu} &=  \omega^{(0)}_{ab\mu}+ K_{\mu ab},\nonumber\\
 K_{\mu ab}&= -\frac{i}{4} \left[ \bar \psi_\mu \gamma_{[a} \psi_{b]} + \bar \psi_a \gamma_\mu \psi_b\right].
 \end{align}
 With the above, Eq. (\ref{deltagAB}) shows that the use of Eq.(\ref{no-k}) leads to the following 
 transformation laws for the fields
 \begin{align}
 \delta e_{a\mu} &= i\bar \psi_\mu \gamma_a \lambda, \nonumber\\
 \delta g_{\mu\nu}&= i\bar \psi_{(\mu} \gamma_{\nu)}\lambda\nonumber\\
 \frac{1}{2} \delta \psi_\mu(x)& = D_\mu \lambda \equiv (\partial_\mu- \Gamma_\mu) \lambda\,.
 \end{align}
 The above are indeed standard supergravity transformations. 
 Next, we exhibit the $k$-dependent parts in the relevant order in the transformation $\xi^\mu$ and
 $\xi^\alpha$ and in $\Delta g_{\mu\nu}, \Delta g_{\mu \alpha}, \Delta_{\alpha\beta}$. Here, one gets
 \begin{align} 
 \Delta \xi_{\mu}& = \frac{ik}{2} (\bar \lambda \gamma^a\theta) (\bar \theta D_{[\mu } \psi_{a]} 
 +\frac{k}{8} (\bar \theta \sigma^{ab}\lambda) (\bar \theta \gamma_\mu D_{[a} \psi_{b]}), \nonumber\\ 
 \noindent
 \Delta \xi^{\alpha}=& -\frac{ik}{8} (\sigma^{ab} \theta)^{\alpha} \bar \lambda D_{[a}\psi_{b]} 
+ \frac{ik}{8} D_{[a} \psi_{b]}^{\alpha} (\bar \theta \sigma^{ab} \lambda) \nonumber\\
& +\frac{ik}{32} \epsilon^{abcd} (\gamma^5 \gamma_d \theta)^{\alpha} \bar \lambda \gamma_c
 D_{[a}\psi_{b]}\,.
 \label{xi-k}
 \end{align}
 Next, we include the k-dependent terms in the metric consistent with the transformations of 
 Eq.(\ref{xi-k}). Here we find the following. 
 
  \begin{align}
  \Delta g_{\mu\nu} &=\frac{k}{4}\bar \psi_{\mu}\psi_\nu+
  \frac{k}{2} \bar \psi_{(\mu} \Gamma_{\nu)} \theta- k \bar\theta \Gamma_\mu \Gamma_\nu \theta\,,\nonumber\\
  \Delta g_{\mu\alpha}&= \frac{k}{2} \bar \psi_{\mu\alpha} - k (\bar \theta \Gamma_\mu)_\alpha 
  -\frac{ik}{2} (\bar\theta \gamma^a)_{\alpha} \bar\theta D_{[\mu} \psi_{a]} \nonumber\\
   & + \frac{k}{4} (\bar\theta \sigma^{ab})_\alpha \bar \psi_\mu D_{[a} \bar \psi_{b]} 
  + \frac{ik^2}{16} (\bar \psi_\mu \sigma^{ab})_\alpha \bar \theta D_{[a}\psi_{b]} \\
&+ \frac{ik^2}{16}(\bar\theta \sigma^{ab})_\alpha 
\bar\psi_\mu D_{[a}\psi_{b]}  
  -\frac{ik^2}{64} \epsilon^{abcd} (\bar \psi_\mu \gamma^5 \gamma^d)_\alpha \bar \theta \gamma_c 
  D_{[a}\psi_{b]}\,, \nonumber\\
    \Delta g_{\alpha\beta} &= -\frac{ik^2}{8} D_{[a} \bar\psi_{b] ~[\alpha}(\bar\theta \sigma^{ab})_{\beta]}   
    -\frac{ik^2}{16} \epsilon^{abcd} (\eta \gamma^5 \gamma_d)_{\alpha\beta} \bar\theta \gamma_c 
    D_{[a} \psi_{b]}\,.
  \end{align} 
    Next we discuss how gauge supersymmetry gives rise to the standard supergravity field equations. 
   They arise from the $k\to 0$ limit of $R_{AB}=\lambda g_{AB}$ of  gauge supersymmetry
   when we keep only the spin 2, spin 3/2 multiplets in the metric. Thus,  the supergravity equations 
   for $g_{\mu\nu}$ and for $\psi^\mu$ arise from the $\theta$ independent components of 
   $R_{\mu\nu}$ and $R_{\mu\alpha}$. Note that the dynamical equations of gauge supersymmetry involve the inverse metric $g^{\alpha\beta}$ which involves an inverse power of $k$, i.e., 
   $g^{\alpha\beta}=\frac{1}{k} \eta^{\alpha\beta} + \cdots$. That is the reason for keeping up to
   ${\cal O}(k^2)$ terms in expansion of the metric. Then it is seen that in the calculation of $R_{AB}$,
 we find terms which are $O(k^{-2})$ and $O(k^{-1})$. However, these terms are canceled internally and
 we move to a smooth limit as $k\to 0$. Thus, in the limit $k\to 0$,   
    the $\theta$ independent part of $R_{\alpha\mu}-\lambda g_{\alpha\mu}$ gives   
   \begin{align}
   R_{\alpha\mu}-\lambda g_{\alpha\mu} &= \frac{3}{4}\bar M_{\mu\alpha} + \frac{1}{16}
   (\bar M^b \gamma_b \gamma_\mu)_\alpha\,,\nonumber\\
   M^\mu& \equiv \epsilon^{\mu bcd}\gamma^5 \gamma_d D_{[c}\psi_{b]}.
 \end{align} 
The above analysis then implies that $  R_{\alpha\mu}=\lambda g_{\alpha\mu}$ leads to $M^\mu=0$,
i.e.,
  \begin{align}
&\epsilon^{\mu bcd}\gamma^5 \gamma_d D_{[c}\psi_{b]} =0.\nonumber\\
&D_\mu= \partial_\mu- \Gamma_\mu,  ~~ \Gamma_\mu= \frac{i}{4} \sigma^{ab} \omega_{ab\mu}
\nonumber\\
& \omega_{ab\mu}= \omega^{(0)}_{ab\mu} + K_{\mu ab}\nonumber\\
&K_{\mu ab}= -\frac{i}{4} \left[ \bar \psi_\mu \gamma_{[a} \psi_{b]} + \bar \psi_a \gamma_\mu\psi_b\right]\,.
 \end{align} 
 Further, the equation $R_{\mu\nu}= \lambda g_{\mu\nu}$ in the $k\to 0$ limit yields the standard 
 supergravity equation for $g_{\mu\nu}$ coupled with the Rarita-Schwinger field $\psi^\mu$. 
 In the above we see that the correct supergravity transformation equations as well as the correct $g_{\Lambda \Pi}$  up to   ${\cal{O}}(\theta^2)$ arise in the $k\to 0$ limit of gauge supersymmetry.
  It is to be noted that the integration of ~Eq.(\ref{deltagAB})
 requires the use of on-shell constraints beyond linear order in $\theta$ which means that one
 needs to use field equations  to complete the integration. Furthermore, integration without the use of field equations,
 that is, off the mass-shell, requires auxiliary   fields~\cite{Breitenlohner:1976nv,Sohnius:1981tp}
which allows gauge completion without the use of field equations \cite{Arnowitt:1978jq}. We note here that the geometry contraction of gage supersymmetry discussed here, which reduces the $OSp(3,1|4N)$ geometry in superspace
to the superspace geometry for $O(3,1)\times O(N)$ with particle content of standard supergravity, may be viewed 
as the counterpart of Inonu-Wigner\cite{Inonu:1953sp} group contraction. However, in spirit they are the same viewed from the end point of contraction, i e.,addition the geometry contraction also includes the dynamics related to the contraction.
 

\section{Appendix B: SUSY and SUSY GUTs}
In this section we expand on the discussion of section 2 and discuss further properties of globally supersymmetric theories. Thus an important result regarding supersymmetric quantum field theory is the existence of non-renormalization theorems\cite{Iliopoulos:1974zv,Grisaru:1979wc,Witten:1981nf,
Kaul:1981wp}. The implication of the theorems is that the superpotential $W(\Phi)$ does not receive perturbative radiative corrections, while the Kähler potential and gauge kinetic terms receive radiative corrections and  get renormalized. Regarding the superpotential, no new terms can be generated radiatively. This further implies that if SUSY is not broken at the tree level, it remains unbroken at the loop level. Thus, cancelation of the quadratic divergence in a supersymmetric theory is mandated by the underlying supersymmetry. Although the superpotential is protected perturbatively, 
 corrections can be generated by nonperturbative effects such as instantons and gaugino condensation\cite{Seiberg:1993vc}. The supersymmetric 
 extension of
 the standard model (MSSM) is based on several earlier works, specifically those of \cite{Fayet:1975yi,Fayet:1976et}.
Before discussion of supersymmetric grand unification (SUSY GUTs), it is informative to discuss non-supersymmetric GUTs. The first step towards unification beyond the standard model of the electroweak and strong interactions was taken by Pati and Salam in 1973\cite{Pati:1973uk,Pati:1974yy} by the formulation of the $G(4,2,2)\equiv SU(4)\times SU(2)_L\times SU(2)_R$
 model where the lepton number was identified as the fourth color by 
 extending the color group so that the left handed color multiplet $(d_r, d_g, d_b)_L$ is extended to 
 read  $(d_r, d_g, d_b,e)_L$ and $(u_r, u_g, u_b)_L$ is extended to read
 $(u_r, u_g, u_b,\nu)_L$. The Pati-Salam model is a left-right symmetric model (for further work along this line, see\cite{Mohapatra:1974gc}). The $G(4,2,2)$ breaks down on a high scale so that $SU(4)\times SU(2)_L\times SU(2)_R \to SU(3)_C\times SU(2)_L\times U(1)_Y$.
 However, because of the product nature of the group, the model is not a complete GUT model. The first complete 
 GUT model is due to Georgi and Glashow\cite{Georgi:1974sy}  in 1974 who proposed $SU(5)$ as the grand unification group whose representations $\bar 5+ 10$ contain all the quarks and leptons.
  The $SO(10)$ grand unification model was proposed in 1975 by 
 Fritzsch and Minkowski, and by Georgi \cite{Fritzsch:1974nn,Georgi:1975_SO10}, which has the additional nice features that a 16-plet
 representation of the group contains a full generation of quarks and leptons.
 In addition, $SO(10)$ can break into the subgroup $SU(4)\times SU(2)\times
 SU(2)$ or $SU(5)\times U(1)$. The GUT groups were subsequently
 extended to include supersymmetry. Thus, $SU(5)$ was extended to
 SUSY $SU(5)$ in\cite{Dimopoulos:1981zb}. Unlike SUSY SU(5) which is essentially unique, SO(10) models are not, as there are several possibilities for the choice
 of the Higgs sector. Further, models based on $SU(5)$ and $SO(10)$
 have the well known doublet-triplet problem in the Higgs sector, which 
 requires that  for the SUSY case a pair of  Higgs doublets be light while
 all the Higgs triplets become heavy. This requires an extra step in the
 grand unified theory to manufacture a mechanism to achieve it. One
 approach is the missing partner
  mechanism \cite{Masiero:1982fe,Grinstein:1982um,Babu:2006nf,Babu:2011tw}
    which, however, requires additional Higgs representations beyond those needed for breaking the GUT gauge group to the standard model gauge group.


\section{Appendix C: Heavy fields and gauge hierarchy in SUGRA GUTs}
The gauge hierarchy relates to the stability of the electroweak scale against large quantum loop corrections. For scalar fields an ordinary gauge theory receives quadratically divergent corrections proportional to the cutoff scale which destabilizes the gauge hierarchy\cite{Gildener:1976ih,Weinberg:1976dk,Weinberg:1978ym,Susskind:1978ms}.  If this  cutoff scale is the grand unification scale or the Planck scale, the loop corrections to the mass $\delta m^2_{\phi}\sim 10^{32}- 10^{36}$ GeV$^2$. In supersymmetry, the quadratic divergences cancel exactly. The question then is
 what happens when supersymmetry is not exact but broken. Here, it
 is shown that the cancelation of the quadratic divergences still holds, provided that the 
 supersymmetry is broken softly. In this case, the corrections to the sparticle masses
 are proportional to the SUSY-breaking scale, i.e., 
$\delta m^2 \sim \mathcal{O}(m_{\text{soft}}^2)$    
\cite{Witten:1981,Kaul:1981wp}.
A full classification of soft terms is given in ref. \cite{Girardello:1982pd}. We note in passing that an
efficient superfield formalism has been developed in \cite{Grisaru:1979wc} using
improved methods for supergraphs for the computation of supersymmetric loops.\\

  Here we further elaborate on how SUGRA GUT is protected from the 
  soft terms of Eq.(\ref{msmG}). Such terms are absent if the hidden sector 
  depends only on the combination $\bar Z \equiv \kappa Z$.
  However, acceptable soft terms can still arise even when the hidden sector field has couplings with the visible sector with the condition that such couplings are suppressed by factors of $\kappa$. The desired suppression can be achieved if we  impose the constraints of Eq.(\ref{alphai}) 
  on the superpotential.
These constraints are sufficient to protect effective low energy theory from undesirable large corrections of Eq.(\ref{msmG}) and are non-trivial. An explicit exhibition of it is given in \cite{Chamseddine:1982jx} for a specific model and a more general proof appears see\cite{Nath:1983aw}.
 There, it is shown that the set of terms of the form ~$m_s M_G (\kappa M_G)^k$ are absent or cancel under the constraints of Eq.(\ref{alphai}). Here we give a brief review of how this comes about since this result forms the basis of setting up  SUGRA GUT.
  In order to discuss dynamics of how Eq.(\ref{alphai}) we first focus on the super Higgs field
  $Z$ and the heavy fields $Z_i$ which satisfy the field equations
 
\begin{align}
\frac{\partial V}{\partial Z}=0,  ~~ \frac{\partial V}{\partial Z_i} =0,
\label{A1}
\end{align}
 These equations are utilized to solve for $Z$ and $Z_i$ in terms of the light fields $Z_\alpha$.
 To this end we expand the VEVs of $Z$, $Z_i$ and $Z_\alpha$ in powers of $\kappa$ so that 

  \begin{align}
  Z= Z^{(-1)} + Z^{(0)} + \cdots; \quad 
  Z_{i}= Z_{i}^{(0)} + Z_{i}^{(1)} + Z_{i}^{(2)} + \cdots; \quad
  Z_{\alpha}= Z_{\alpha}^{(1)} + Z_{\alpha}^{(2)}  + \dots,
  \end{align}
  where $Z^{(n)}$ is of order $\kappa^n$. Our aim is to show that the effective low energy 
  potential $V_{eff}$ is independent of $M_G$, i.e., 
  \begin{align}
  V_{eff}(Z_{\alpha})= V[ Z_i(Z_{\alpha}); ~Z_\alpha;  ~Z(Z_\alpha)],
  \label{A2}
  \end{align}
    is independent of the scale $M_G$. The analysis is rather intricate. We can start by writing the equations of motion for the fields
$Z_A$ which take the form 

\begin{align}
T_{AB} G_{B}=0,
\label{A3}
\end{align}
where $T_{AB}$ is given by
\begin{align}
 T_{AB} = W_{,AB} + \frac{1}{2} \kappa^2 (Z_A G_B + Z_B G_A) -\frac{1}{4} \kappa^4 Z_A Z_B W -\kappa^2 \delta_{AB} W,
 \label{A5}
\end{align}
and where  $W_{,A}\equiv \partial W/\partial Z_A$.Further, 
  $G_A$ is given by
\begin{align}
G_A= W_{,A} + \frac{1}{2} \kappa^2 Z_A^{\dagger} W.
\label{A4}
\end{align}
For further analysis, it is useful to define rescaled field variables $z_\alpha, z_i, z$ 
as follows 
\begin{align}
Z_{\alpha}= m_s z_\alpha, ~~ Z_i= M_G z_i, ~~ Z=M_{Pl} z.
\label{A7}
\end{align}
Additionally, it is convenient to define dimensionless quantities $\overline{W}$, $\overline{G}_\alpha$,
$\overline{G}_i$ and $\overline{G}_Z$ by rescaling as follows:
\begin{align}
W= m^2 M_{Pl} \overline{W}, ~
~G_{\alpha} = m_s^2 \overline{G}_{\alpha}, ~G_{i} = m_s^2 \overline{G}_{i}, ~~G_Z= m^2 \overline{G}_Z.
\label{A8}
\end{align}
Using  Eq.(\ref{A3}), one can then obtain equations when 
 $A=Z, Z_i, Z_{\alpha}$. 
 Thus the field equation for the super Higgs field $Z$ takes the form 
    \begin{align}
\left [\frac{1}{m_s} {W}_{SH,ZZ} + (z \bar G_Z -{\overline W} - \frac{1}{4} z^2 \bar W)
 + \frac{1}{2} (\epsilon \delta_s  z_i  {\overline G}_i + \delta_s^2 z_{\alpha} {\overline G}_{\alpha})\right ] {\overline G}_Z\nonumber\\
 +\frac{1}{4} \epsilon \delta_s z z_i  {\overline G}_i {\overline W}
 + \frac{1}{2} \delta_S^2 (z {\overline G}_i^2 + z {\overline G}_{\alpha}^2 - \frac{1}{2} z z_{\alpha}
 {\overline G}_{\alpha} {\overline W}) =0,
 \label{gzz}
 \end{align}
and the field equation for the heavy fields $Z_i$ takes the form of 
      \begin{align}
 [W_{,ij} +\frac{M_G}{2} \left\{\delta_s^2 (z_i\bar G_j + z_j \bar G_i)  -\frac{1}{2} \epsilon \delta_s z_iz_j \bar W
 \right\}
 + m_s \delta_{ij} \left\{\frac{1}{2} z \bar G_z - \bar W + \frac{1}{2} \delta_s^2 z_{\alpha} \bar G_{\alpha}
 \right\}] {\overline G}_j   \nonumber\\
  +  M_G \left[ \frac{1}{2} z_i {\overline G}^2_Z
   -\frac{1}{4} z_i z \overline{W}  {\overline G}_Z
   + \frac{1}{2} \delta_s^2 z_i ({\overline G}_{\alpha}^2 - \frac{1}{2} {\overline G}_{\alpha} \overline{ W})\right]=0,
   \label{A13a}
 \end{align}

while the field equation for the light fields is given by
  \begin{align}
 \left (\frac{1}{m_s} {W}_{,\alpha \beta} + \delta_{\alpha \beta} (\frac{1}{2} z \overline G_Z - \overline{W})
    + \frac{1}{2} \epsilon \delta_s \delta_{\alpha\beta} z_i \overline{G}_i
  + \frac{1}{2} \delta_s^2 (z_\alpha \overline{G}_\beta + z_\beta \overline{G}_\alpha)
  -\frac{1}{4} z_\alpha z_\beta \overline{W} \right ) \overline{G}_\beta
  \nonumber\\
  + \left (\overline{W}_{,\alpha i} \overline{G}_i + \frac{1}{2} z_{\alpha} \overline{G}_Z^2 -\frac{1}{4} z_\alpha z
  \overline{W}~\overline{G}_z \right)
  -\frac{1}{4} \epsilon \delta_s z_\alpha z_i  \overline{W} ~\overline{G}_i + \frac{1}{2} \delta_s^2 z_\alpha
  \overline{G}_i^2 =0.~~~~~~~~~~~
  \label{A10}
   \end{align}
   The field equations 
Eqs.(\ref{gzz}, \ref{A13a}, \ref{A10}) contain two parameters of  smallness

 \begin{align}
 \epsilon = \kappa M_G  \sim 10^{-2}; ~~~\delta_s= \kappa m_s \sim 10^{-16}.
 \end{align}
 Since $W_{,ij}$ is proportional to the heavy sector mass matrix, i.e., $W_{,ij} \sim M_G$, we see  that
Eq.(\ref{alphai})  allows solutions of Eqs.(\ref{gzz}, \ref{A13a}, \ref{A10})
 where all the dimensionless  barred and lower case
fields are $O(1)$ (with  corrections of size $\epsilon \delta_s$ and $\delta_s^2$). This is remarkable
since on dimensional grounds one would have expected $G_i\sim M_G^2$.
Eqs.(\ref{gzz}) and (\ref{A13a}) allow one to determine $Z$ and $Z_i$ in terms of $Z_{\alpha}$ and when
inserted into Eq.(\ref{A10}) gives the equation to determine $z_{\alpha}$ which is then independent of
$M_G$
(aside from small $\epsilon \delta_s, \delta^2_s$ corrections). After a detailed analysis, it is  shown that the resultant Eq.(\ref{A10}) is the field equation deducible
from varying an effective potential $V_{eff}$ depending only on low energy fields 
$Z_{\alpha}$\cite{Nath:1983aw}.
From the analysis above we make the  observation that the low energy field equation 
in terms of the light fields while free of large masses $M_G$ and $M_{Pl}$ is not totally
independent of them. However, they enter only in the combination $\epsilon \delta_s$ and 
$\delta^2_s$ and thus make essentially very small contributions. 
We now discuss the implications of the constraints of Eq.(\ref{alphai}) on the couplings 
allowed in the superpotential. Thus, consider the coupling  $y_{i\alpha\beta} Z_iZ_{\alpha} Z_{\beta}$. This coupling gives $W_{,i\alpha }$ that depends on $Z_{\beta}$ with a light 
VEV of size $O(m_s)$, and thus the first condition of Eq.(\ref{alphai}) is satisfied. 
The second condition of Eq.(\ref{alphai}) is satisfied if the VEV of $Z_i$ vanishes.
However, it is later shown that such heavy-heavy-light couplings destabilize 
the gauge hierarchy
if $Z_\alpha$ couples to other light fields ~\cite{Nilles:1982mp,nsw,sen}.)
Also
\cite{AHCThesis}.

\bibliographystyle{unsrt}
\bibliography{sample}

\end{document}